\documentclass[11pt]{article}
\usepackage{amssymb,amsmath}
\usepackage{bm} 
\usepackage{booktabs} 
\usepackage{array}
\usepackage{latexsym}
\usepackage{graphicx}
\usepackage{color}
\usepackage{datetime}
\usepackage[nosort]{cite}
\usepackage{verbatim}
\usepackage{enumerate}
\usepackage{chngpage} 
\usepackage{mathrsfs}
\usepackage{euscript}
\usepackage{psfrag}
\usepackage{physics}
\usepackage{tensor}
\usepackage{dsfont}

\usepackage{mciteplus}

\usepackage[colorlinks=true,      linkcolor=blue,      urlcolor=blue,      
            filecolor=blue,      citecolor=blue,       pdfstartview=FitH,     
						pdfpagemode=UseNone,      bookmarksopen=true]{hyperref}  
\usepackage[all]{hypcap}     

\numberwithin{equation}{section} 

\definecolor{cardinal}{rgb}{0.6,0,0}
\definecolor{darkgreen}{rgb}{0,0.4,0}
\definecolor{golden}{rgb}{0.92, 0.7, 0}
\definecolor{midnight}{rgb}{0, 0, 0.5}
\definecolor{darkblue}{rgb}{0, 0, 0.7}
\definecolor{purple}{rgb}{0.5, 0, 0.5}

\def\oneone{\mathds{1}}
\def\IC{\mathbb{C}}
\def\Neql#1{{\cal N}\!=\!{#1}}
\def\coeff#1#2{\relax{\textstyle {#1 \over #2}}\displaystyle}

\def\IR{\mathbb{R}}

\def\IT{\mathbb{T}}

\def\cA{{\cal A}}
\def\cB{{\cal B}}

\def\cD{{\cal D}}

\def\cK{{\cal K}}
\def\cL{{\cal L}}
\def\cM{{\cal M}}
\def\cN{{\cal N}}

\def\cP{{\cal P}}
\def\cQ{{\cal Q}}
\def\cR{{\cal R}}

\def\cV{{\cal V}}

\def\cX{{\cal X}}

\def\scrC{{ \mathscr{C}}}

\def\nBPS#1{$\frac{1}{#1}$-BPS}

\begin{document}

\phantom{AAA}
\vspace{-10mm}

\begin{flushright}
%
%
\end{flushright}

\vspace{1.9cm}

\begin{center}

{\huge {\bf Supersymmetry and Superstrata}}\\
{\huge {\bf \vspace*{.25cm} in Three Dimensions }}

\vspace{1cm}

{\large{\bf {Anthony Houppe$^{1}$  and  Nicholas P. Warner$^{1,2,3}$}}}

\vspace{1cm}

$^1$Institut de Physique Th\'eorique, \\
Universit\'e Paris Saclay, CEA, CNRS,\\
Orme des Merisiers, Gif sur Yvette, 91191 CEDEX, France \\[12 pt]
\centerline{$^2$Department of Physics and Astronomy}
\centerline{and $^3$Department of Mathematics,}
\centerline{University of Southern California,} 
\centerline{Los Angeles, CA 90089, USA}

\vspace{10mm} 
{\footnotesize\upshape\ttfamily anthony.houppe @ ipht.fr, warner @ usc.edu} \\

\vspace{2.2cm}
 
\textsc{Abstract}

\end{center}

\begin{adjustwidth}{3mm}{3mm} 
 
\vspace{-1.2mm}
\noindent
We analyze the supersymmetry transformations of gauged $SO(4)$ supergravity coupled to extra hypermultiplets in three dimensions, and find large families of smooth BPS solutions that preserve four supersymmetries.  These BPS solutions are part of the consistent truncation of some families of  six-dimensional superstrata.  From the three-dimensional perspective,  these solutions give rise to ``smoothly-capped BTZ'' geometries.  We show how the  twisting of the spin connection, the holomorphy of the fields, and the Chern-Simons connections all play an essential role in the existence of these supersymmetric solutions.  This paper also closes the circle on the consistent truncation of superstrata, showing precisely how every feature of the superstratum enters into the three-dimensional BPS structure.
 
\end{adjustwidth}

\thispagestyle{empty}
\newpage


\baselineskip=17pt
\parskip=5pt

\setcounter{tocdepth}{2}
\tableofcontents

\baselineskip=15pt
\parskip=3pt

\section{Introduction}
\label{sec:Intro}

Superstrata have provided some of the most remarkable and broad families of microstate geometries.  They approximate the exterior regions of black-holes to high precision, and yet cap off smoothly, at arbitrarily high red-shift, yielding smooth, BPS horizonless geometries.  What makes superstrata all the more remarkable is that they have a precisely-known holographic dictionary in which the magnetic fluxes that support the superstrata can be directly related to coherent excitations in the dual conformal field theory. Indeed,   this holographic dictionary has undergone highly non-trivial tests using precision holography in which the structure of the supergravity background has been checked against correlation functions in the CFT. 

To date, the known superstrata are all supersymmetric, or BPS, and one of the current major priorities of the microstate geometry programme is to obtain non-extremal superstrata.  {\it A priori,} this seems an impossible challenge.  Superstrata are constructed in six-dimensional supergravity, and even the simplest of BPS superstrata depend on four variables, and the most general depend on five variables.  It is, therefore, to be expected that non-trivial, non-extremal superstrata will be time-dependent and so will depend on, at least, five variables.  One can, of course, try to find such solutions in perturbation theory, but to go beyond this, and  find more generic, non-extremal superstrata would  seem to be out of reach of even the most advanced numerical methods. 

However, a new family of consistent truncations \cite{Mayerson:2020tcl} has shown how to reduce  a non-trivial ``superstratum'' sector  of  six-dimensional $(1,0)$ supergravity,    to a family of gauged supergravities in three-dimensions.  This provides a promising new approach to the challenge of finding non-extremal superstrata, and may well lead to a breakthrough in the near future.

There are  many examples in which gauged supergravity in low dimensions represent consistent truncations of higher dimensional supergravity theories.  A consistent truncation is much more than having a  low-dimensional effective field theory:  A consistent truncation means that solving the dynamics in the lower dimension provides an {\it exact} solution to the higher dimensional theory.  This means that any dependence of a solution on the compactification manifold is handled through the machinery of the consistent truncation and does not need to be solved directly. A consistent truncation can therefore dramatically simplify the  BPS  equations and, even more importantly,  the equations of motion.  

The ``heavy lifting'' in establishing a consistent truncation is to prove the ``uplift formulae, '' which show how all the fields of the higher dimensional theory can be fully reconstructed from the fields of the lower dimensional theory.  This often means that some extremely complicated solutions of the higher-dimensional supergravity, that generically depend on many variables, can actually be encoded in far simpler solutions, involving fewer variables, in the  lower-dimensional theory.    There is now a forty-year history of studying consistent truncations on sphere compactifications, and slowly, but surely, the uplift formulae have been obtained and proven.  Over the last 20 years, consistent truncation and gauged supergravity methods have found extensive applications in  the analysis of holographic field theory. 

This paper focussed on the consistent truncation of particular $(1,0)$ and $(2,0)$ supergravities in six dimensions that are compactified on $S^3$ down to families of three-dimensional gauged supergravities.  There is an extensive literature on such consistent truncations  but the consistent truncations of six-dimensional $(1,0)$  supergravity {\it coupled to two (or more)  anti-self-dual tensor multiplets}  was only recently established  \cite{Mayerson:2020tcl}.  It is precisely this family of six-dimensional supergravities that can capture the smooth geometries of superstrata: at least two anti-self-dual tensor multiplets are essential to having  smooth geometries, and to capturing the dual description of correlators in the holographic field theory. 

The significance of this consistent truncation is that it shows how particular families of BPS superstrata, depending on four variables in six dimensions, reduce to a three-dimensional solution that depends on only two variables.  Moreover, it is also evident that the three-dimensional theory can encode the collisions of BPS and anti-BPS superstratum waves, and thus have the potential to capture a whole range of non-extremal, non-BPS microstate geometries. As a result, such non-extremal solutions could  be described in terms of functions of, at most, three variables.  Apart from the dramatic simplification afforded by such a truncation, it also brings the problem within striking distance of numerical methods.  As an added bonus, we also know the precise holographic dictionary for such solutions and can therefore study, and thoroughly test,  such a collision of superstratum waves in the  dual CFT. 

Before attempting this ambitious programme on non-extremal  superstrata, it is essential to study the relevant three-dimensional supergravities and lay out all their structural features, and thoroughly test the results of \cite{Mayerson:2020tcl}. In \cite{Mayerson:2020tcl}, it was  shown how two highly-non-trivial  families of superstrata can be reduced to a complete description in terms of gauged supergravity in three dimensions.   The uplift formulae were also obtained in \cite{Mayerson:2020tcl}, and  were extensively tested against the six-dimensional equations of motion using numerical methods.  In this paper we make a detailed analytical  study of the BPS equations and show how they do indeed give rise to  non-trivial families of BPS superstrata. 

The results of \cite{Mayerson:2020tcl}, and the work we will describe here, rests on a remarkable body of earlier work on gauged three-dimensional supergravities and upon the consistent truncations of other six-dimensional supergravities.   A thorough survey of this literature would take us far afield, but in this paper we have drawn heavily upon \cite{Cvetic:2000dm,Cvetic:2000zu, Nicolai:2001sv, Nicolai:2001ac, Nicolai:2003bp, Nicolai:2003ux,Deger:2014ofa,Samtleben:2019zrh}. The results in 
\cite{Deger:2014ofa,Samtleben:2019zrh} describe the consistent truncation of $(1,0)$ supergravity coupled to only one anti-self-dual tensor multiplet and so this was an immensely valuable guide to the construction in \cite{Mayerson:2020tcl}\footnote{The work in \cite{Mayerson:2020tcl} made a lot of non-trivial tests of the uplift formulae of \cite{Samtleben:2019zrh} and this led to some minor corrections and clarifications.}.  Here we are focussing on the supersymmetries in three dimensions, and for this we make extensive use of \cite{Nicolai:2001sv,Nicolai:2001ac, Nicolai:2003bp, Nicolai:2003ux}.  

There is about a 15-year gap between the work on supersymmetry in three dimensions and the more recent work on consistent truncation. In that time there has been a major evolution in the construction, expression and conventions underpinning the results in three and six dimensions. One of the other purposes of this paper is to ``bridge the 15-year gulf,'' and provide a direct translation between these two eras in supergravity.  It turns out that there are many and varied, explicit and implicit, choices of convention in the three or four primary sources for this paper.  Part of our task will be to lay out every aspect of this as a foundation for future work.   In what we hope will be a useful public service, we will carefully catalogue all the essential details. 

Quite apart from the motivation of superstrata, the paper is also an analysis of BPS solutions in a particular three-dimensional gauged supergravity theory. Some supersymmetric solutions of these supergravity theories have been found and analyzed.  However, to date, such BPS solutions have typically been the simple AdS$_3$ vacua or singular black strings \cite{Deger:2019jtl}.  From the perspective of three-dimensional supergravity, the BPS solutions we present here are remarkable: they represent extremely rich families  of smooth, capped BTZ geometries.   This  should be no surprise because they are superstrata, but from the three-dimensional perspective they represent a rather dramatic evolution.  

It is also very interesting to see everything that goes into the three-dimensional BPS solutions. Indeed they seem to exploit the full range of options: (i) Simple projections of the supersymmetries of the ``parent theory,'' (ii) holomorphy of the scalar fields; (iii) twisting spin connections with gauge fields and (iv) highly non-trivial use of the Chern-Simons structure.  

We will begin in Section \ref{Sect:3Dtheory} with a review of the relevant three-dimensional gauged supergravities.  We start with the $\Neql8$ theory, with $16$ supersymmetries, that underlies the D1-D5 system at its ``self-dual'' point.  We do this because the details of the supersymmetries of this model were established 20 years ago in \cite{Nicolai:2001ac}.  We then discuss the truncation to the $\Neql4$ theory, with eight supersymmetries that underlies the more recent work  \cite{Deger:2014ofa,Samtleben:2019zrh,Mayerson:2020tcl}  of more direct relevance to superstrata.   We then catalogue all the essential parts of the action.  For those in a hurry, and for later convenience, we have provided ``executive summaries'' of the action in Section \ref{sec:3Daction} and of the BPS equations in Section \ref{sec:BPS}.

In Section \ref{sec:superstrata_in_3_dimensions} we first summarize the results of \cite{Mayerson:2020tcl} that give the three-dimensional description of the ``$(1,m,n)$ superstrata.'' We then   use this data  to compute all the details of  the three-dimensional fields that appear in the BPS equations.  We also make some gauge transformations that simplify the general solution found in  \cite{Mayerson:2020tcl}.  

Section \ref{sec:Supersymmetry} contains a summary of  the  conditions that we need to impose on the spinors and the fields so as to solve the BPS equations.  We do indeed find that the three-dimensional solutions have four supersymmetries as do the original superstrata.  Section \ref{sec:10n_supersymmetry} contains 
 an analysis of the BPS equations for the smaller class of $(1,0,n)$ superstrata in three dimensions and in Section \ref{sec:1mn_supersymmetry} we perform the BPS analysis for the three-dimensional formulation of the generic  $(1,m,n)$ superstrata.  As we will see in  Section \ref{sec:1mn_supersymmetry}, the analysis of the simpler superstrata  in Section \ref{sec:10n_supersymmetry} plays an essential role in the more general analysis.   
 
 Finally, we makes some concluding remarks in  Section \ref{sec:Conclusions}, and some technical details have been put in an Appendix.

\section{Three-dimensional supergravity theories}
\label{Sect:3Dtheory}

Our purpose here is to give a full and complete description of the three-dimensional gauged supergravity and its supersymmetries.  It is also to provide a translation between the supergravity theory in which the supersymmetries were analyzed  \cite{Deger:1999st,Nicolai:2001ac} and the more recent discussions that involve the purely bosonic actions \cite{Deger:2014ofa,Samtleben:2019zrh,Mayerson:2020tcl} for which the uplift formulae have been derived.   All of these references use different conventions and, in testing the supersymmetry, a significant effort  goes into making the translation between the various formulations.  We will therefore try to spell out many of the explicit details.

Initially, we will work with the $\Neql8$ ($16$ supersymmetries) theory that was analyzed in  \cite{Nicolai:2001ac}. The theory we ultimately seek to analyze is the  $\Neql4$ ($8$ supersymmetries)  theory that underlies the D1-D5 system and that was used in \cite{Deger:2014ofa,Samtleben:2019zrh,Mayerson:2020tcl}. We will show how this theory is a truncation of the larger theory of \cite{Nicolai:2001ac}.

\subsection{A summary of notation and conventions}
\label{sec:notation}

\subsubsection{Group theory}
\label{sec:groups}

We are going to be working with the group $G = SO(8,n)$, and its subgroup $H = SO(8) \times SO(n)$.     Following \cite{Nicolai:2001ac}, we   use calligraphic indices as adjoint labels of $G = SO(8,n)$.  We  use barred, capital Latin indices, $\bar I, \bar J, \dots$ to denote the vector of $SO(8,n)$, unbarred capital Latin indices, $I, J, \dots$,  to denote the vector of $SO(8)$, and small Latin indices, $r, s, \dots$ to denote the vector of $SO(n)$.  In the standard way, the adjoint indices of $G = SO(8,n)$,  $SO(8)$ and $SO(n)$ can be written as skew pairs $\bar I  \bar J$,  $IJ$ and $rs$. Such a labelling double counts the adjoint but when we use this notation we will always sum over all indices without any implicit factors of $\frac{1}{2}$.  Later in the discussion we will restrict to $G = SO(4,n)$, and its subgroup $H = SO(4) \times SO(n)$, and this will result in the obvious restrictions on the index ranges.  

We define the $G = SO(8,n)$ invariant matrix in its canonical form:
\begin{equation}
\eta^{\bar I \bar J} ~=~ \eta_{\bar I \bar J}~\equiv~ {\rm diag} \big(1,1,\dots, 1, -1,-1,\dots -1 \big) \,, \qquad \eta^{IJ} ~=~ \eta_{IJ} ~=~ \delta^{IJ} \,, \qquad \eta^{rs} ~=~ \eta_{rs} ~=~ -\delta^{rs} 
\label{invmat1}
\end{equation}
Later, when we restrict to $G = SO(4,n)$, we will introduce $\hat \eta^{\bar I \bar J} = \hat \eta_{\bar I \bar J}$, which will be adapted to the $GL(4,\IR)$ basis  (see (\ref{invmat2})).

The matrices, $t^\cM$, will denote generators of the adjoint of $G$.  In the obvious manner, it will be convenient to define
\begin{equation}
\Big \{ t^\cM \Big\}~\equiv~ \Big\{ X^{\bar I \bar J} = - X^{\bar J \bar I}\Big\}~\equiv~ \Big\{X^{IJ} = - X^{JI} , X^{rs} = -X^{sr},Y^{Ir} =- Y^{rI}  \Big\}
\label{genbasis}
\end{equation}
The structure constants are defined, as usual, via $\big [\, t^\cM  \,, \, t^\cN \, \big ] =f^{\cM \cN}{}_\cP \, t^\cP$, which we write as
\begin{equation}
\begin{aligned}
\Big [\,  X^{\bar I \bar J}   \,,\,  X^{\bar K \bar L}  \, \Big ] ~=~ & f^{{\bar I \bar J}  \, {\bar K \bar L} }{}_{\bar M \bar N}  \,  X^{\bar M  \bar N}  \\
~=~ &   - \eta^{\bar I \bar K} X^{\bar J \bar L}~+~ \eta^{\bar I \bar L} X^{\bar J\bar K  } ~+~\eta^{\bar J \bar K} X^{\bar I \bar L}~-~ \eta^{\bar J \bar L} X^{ \bar I \bar K}\,.
\end{aligned}
\label{comms1}
\end{equation}
which leads to:
\begin{equation}
\begin{aligned}
f^{\bar I\bar J \, \bar K\bar L}{}_{\bar M\bar N}  ~=~ &
-  \coeff{1}{2}\, \eta^{\bar I \bar K} \, \big( \delta^{\bar J}_{\bar M} \,\delta^{\bar L}_{\bar N}   - \delta^{\bar J}_{\bar N} \,\delta^{\bar L}_{\bar M}\big)~+~  \coeff{1}{2}\,\eta^{\bar I \bar L} \, \big( \delta^{\bar J}_{\bar M} \,\delta^{\bar K}_{\bar N}   - \delta^{\bar J}_{\bar N} \,\delta^{\bar K}_{\bar M}\big) \\
  & ~+~ \coeff{1}{2}\,\eta^{\bar J \bar K} \, \big( \delta^{\bar I}_{\bar M} \,\delta^{\bar L}_{\bar N}   - \delta^{\bar I}_{\bar N} \,\delta^{\bar L}_{\bar M}\big) ~-~  \coeff{1}{2}\, \eta^{\bar J \bar L} \, \big( \delta^{\bar I}_{\bar M} \,\delta^{\bar K}_{\bar N}   - \delta^{\bar I}_{\bar N} \,\delta^{\bar K}_{\bar M}\big) \,.
\end{aligned}
\label{structureconsts}
\end{equation}
Note that the factors of $\frac{1}{2}$ appear because  we are summing over all values of ${\bar M\bar N}$ and so this leads to a double counting of the generators.    
 
In terms of explicit matrix representations, (\ref{comms1}) and  (\ref{structureconsts}) correspond to using the matrix generators: 
\begin{equation}
\big( \,X^{\bar I \bar J} \, \big)_{\bar K}{}^{\bar L}  ~=~   \delta^{\bar I}_{\bar K}\,  \eta^{\bar J \bar L} ~-~  \delta^{\bar J}_{\bar K}\,  \eta^{\bar I \bar L}  \,.
\label{matform}
\end{equation}
One should note that these choices are the same as the conventions used in equation (A.3) of \cite{Nicolai:2001sv}\footnote{ However, in this reference, equation (A.3) is inconsistent with (A.1)!}, and equation (2.4) in \cite{Samtleben:2019zrh}\footnote{However, as noted in  \cite{Mayerson:2020tcl}, there are inconsistencies in the  gauge action of   \cite{Samtleben:2019zrh}, and these suggest an inconsistent usage of the structure constants or gauge matrices. The gauge action in  \cite{Mayerson:2020tcl} is correct and consistent.} but have the opposite signs to those of \cite{Mayerson:2020tcl},  equations (2.13) and (2.14).  The conventions we use here  appear  to be the use used in  \cite{Nicolai:2001ac}, and match those of equation (3.9) of \cite{Nicolai:2003ux}\footnote{This requires a small correction explained below.}.   While the gauge matrices and structure constants that we use here have the opposite sign to those of \cite{Mayerson:2020tcl}, we will eventually arrive at the same formulation as \cite{Mayerson:2020tcl} through the choice of the sign of a gauge coupling, or the ultimate sign of the embedding tensor.  As we will see, the signs of these generators are crucial to showing that the supersymmetric solutions in \cite{Mayerson:2020tcl} are indeed consistent with the supersymmetry variations of  \cite{Nicolai:2001ac}. This  provides many non-trivial tests of all the details we are cataloging here\footnote{The results in this paper therefore provide detailed  confirmation (and the occasional correction or clarification) of the conventions and results in the literature over the last 20 years.}.

While it is not directly relevant to our discussion here, to match our conventions to those of the commutators in equation (3.9) of \cite{Nicolai:2003ux} one must reverse the signs of the generators, $X^{rs}$.  This sign reversal is natural because the metric (\ref{invmat1}) is negative definite on $SO(n)$ and this passes into  (\ref{comms1}) and (\ref{matform}):  Reversing the signs of the $X^{rs}$  give them commutators for a positive definite metric on $SO(n)$.  Our computations are not sensitive to this sign and we will stay with the conventions above.

\subsubsection{$SO(8)$ spinors}
\label{sec:SO8-conventions}

The $\Neql8$ theory  has an $SO(8)$ $\cR$-symmetry, and the fermions transform in the spinor representations. Thus we will need $16 \times 16$, $SO(8)$ $\Gamma$-matrices that satisfy:
\begin{equation}
\big\{ \, \Gamma^I\,, \,  \Gamma^J \, \big\} ~=~ 2\,\delta^{IJ}\, \oneone_{16 \times 16} \,.
\label{anticomms}
\end{equation}
We use capital Latin indices $I,J,K, \dots $ to denote vector indices,  capital Latin indices $A,B, C, \dots $ to denote spinors in the $8^+$ Weyl  representation and dotted capital Latin indices $\dot A,\dot B, \dot C, \dots $ to denote spinors in the $8^-$ Weyl  representation.    We use a representation of the $\Gamma$-matrices where they are real and symmetric and in which the non-trivial, $8 \times 8$ blocks are the off-diagonal pieces: $\Gamma^J_{A \dot A}$ and  $\Gamma^J_{ \dot A A}$.  The helicity projector is:
\begin{equation}
(\Gamma^{12345678})_{AB}  ~=~ \delta_{AB} \,, \qquad  (\Gamma^{12345678})_{\dot A \dot B}  ~=~ - \delta_{\dot A \dot B} \,.
\label{helproj1}
\end{equation}
%

\subsubsection{Space-time metric and spinors}
\label{sec:st-conventions}

Much of the literature on three-dimensional gauged supergravity uses the conventions set up in \cite{Marcus:1983hb}, and we will follow suit.
This means that the metric has signature $(+- -)$.  The $2 \times 2$ space-time gamma matrices are:
\begin{equation}
\gamma^0  ~=~ \sigma_2 
 ~=~
\left( \begin{matrix} 
0 & -i \\
i & 0
\end{matrix} \right) \,,
\qquad \gamma^1  ~=~ i \, \sigma_3 
 ~=~
\left( \begin{matrix} 
i &0 \\
0 & -i
\end{matrix} \right) \,,
 \qquad \gamma^3 ~=~ i \, \sigma_1 
 ~=~
\left( \begin{matrix} 
0 & i \\
i & 0
\end{matrix} \right) 
 \,.
\label{st-gamas}
\end{equation}
and we have $\gamma^{012}= - i \oneone_{2 \times 2}$.

The orientation is set by taking (in frames):
\begin{equation}
\epsilon^{012}   ~=~ \epsilon_{012} ~=~ +1 \,.
\label{epsdefn}
\end{equation}
We will use $\epsilon^{abc}$ and $\epsilon_{abc}$ to denote the permutation signature that takes values $0, \pm 1$.  The covariant $\varepsilon$-symbol will be denoted 
\begin{equation}
\varepsilon_{\mu \nu \rho}   ~=~ e \, \epsilon_{\mu \nu \rho}  \,, \qquad \varepsilon^{\mu \nu \rho}   ~=~ e^{-1} \, \epsilon^{\mu \nu \rho}   \,,
\label{varepsdefn}
\end{equation}
where $e = \sqrt{| g|}$ is the frame determinant.

\subsubsection{The three-dimensional metric}
\label{sec:metric}

The most general, three-dimensional metric will depend on  three arbitrary functions: six metric components minus three functions from coordinate transformations.   One way to realize this is to fix two of the coordinate transformations to arrive at a conformally-flat spatial base.  The time direction can also have its own scale factor, and there can also be two-component, angular-momentum vector, $k$.  One can use the coordinate re-definition of $t$ to gauge $k$ so that it only has one spatial component.  We therefore claim that, at least locally, the most general three-dimensional metric can be re-cast in the form:
\begin{equation}
\begin{aligned}
ds_{3}^{2}  ~=~&  \Omega_1^{2} \, (dt  +  k_v \, dv  )^2~-~  \Omega_0^{2}\,\frac{|d\xi|^{2}}{(1-|\xi|^{2} )^{2}}  \\
  ~=~& \Omega_1^{2} \, (dt  +  k_v \, dv  )^2~-~  \Omega_0^{2}\,\bigg(\frac{dr^2}{r^2 + a^2} ~+~\frac{2}{R_y^2 \, a^4 } \,r^2\,(r^2 + a^2) \, dv^2 \bigg)  \,,
\end{aligned}
\label{genmet}
\end{equation}
where 
\begin{equation}
\xi ~\equiv~\frac{r}{\sqrt{r^2+ a^2}} \, e^{i \frac{\sqrt{2} v}{R_y} }\,. 
\label{xidef}
\end{equation}

The three arbitrary functions are $\Omega_0$,  $\Omega_1$ and $k_v$.  We are using the coordinates $(t,r,v)$ because they are well adapted to the discussion of asymptotically AdS$_3$ space, and superstrata.  Indeed, the metric on global AdS$_3$ of radius $R$,  can be written as:
\begin{equation}
\begin{aligned}
ds_{3}^{2}  ~=~ R^2 \, \bigg(  R_y^{-2} \, \big(dt  +  R_y \, \mathscr{A}  \big)^2~-~ \frac{|d\xi|^{2}}{\left(1-|\xi|^{2} \right)^{2}} \bigg) \,,
\end{aligned}
\label{AdSMet1}
\end{equation}
where
\begin{equation}
\mathscr{A} ~\equiv~  \frac{i}{2} \left( \frac{\xi \, d\bar{\xi} - \bar{\xi} \, d\xi}{1-|\xi|^{2}} \right)  \,. 
\label{Aconn1} 
\end{equation}

As usual on AdS$_3$, we interchange between $(t,y)$ coordinates, and null coordinates $(u,v)$ via: 
\begin{equation}
u ~\equiv~   \frac{1}{\sqrt{2}}\, (t-y) \,,  \qquad v ~\equiv~    \frac{1}{\sqrt{2}}\, (t+y)  \,,
 \label{uvdefn}
\end{equation}
The parameter $R_y$, is the radius of the $y$-circle: 
\begin{equation}
y ~\cong~ y \,+\, 2 \pi R_y  \,. 
\label{yperiod}
\end{equation}

We will return to these metrics later, but here we will fix our frame orientations.  We will take frames with 
\begin{equation}
e^0  ~\sim~(dt  +  k_v \, dv  ) \,, \qquad  e^1  ~\sim~ dr  \,,  \qquad  e^2  ~\sim~ dv   \,. 
\label{framedirs}
\end{equation}
and use the $\epsilon$-symbols defined in (\ref{epsdefn}), or (\ref{varepsdefn}).  This means that our volume form has the orientation: 
\begin{equation}
vol_3   ~\sim~ dt \wedge dr \wedge dv  ~\sim~  du \wedge dr \wedge dv   \,. 
\label{volorient}
\end{equation}
The orientation in  \cite{Mayerson:2020tcl} was given as: 
\begin{equation}
e^{-1}\varepsilon_{uvr} ~=~ -\varepsilon \qquad \Rightarrow \qquad vol_3   ~\sim~   -\varepsilon  \, du \wedge dv \wedge dr  ~=~ \varepsilon  \, du \wedge dr \wedge dv   \,, 
\label{eq:3Dorientation}
\end{equation}
where $\varepsilon = \pm 1$ is a parameter introduced in  \cite{Mayerson:2020tcl}.   As we will discuss, we will take $\varepsilon=-1$, and so, on the face of it, we seem to be using different orientations to that of \cite{Mayerson:2020tcl}.   However, one should remember that \cite{Mayerson:2020tcl} uses the opposite of our metric signature and so raising all the indices to create $\varepsilon^{\mu \nu \rho}$ flips the sign in \cite{Mayerson:2020tcl} but does not change the sign here.  Thus  the tensors, $\varepsilon^{\mu \nu \rho}$, have:
\begin{equation}
e \, \varepsilon^{urv} ~=~\epsilon^{012}  ~=~ -  \varepsilon  \,, 
\label{eq:3Dorientation2}
\end{equation}
and are therefore the same as those of \cite{Mayerson:2020tcl}  (once we take  $\varepsilon=-1$).  This is the important convention because it is (\ref{eq:3Dorientation2}) that enters the expressions for the Chern-Simons terms and their kindred.

\subsection{The   $\Neql8$ theory}
\label{sec:Neql8}

Our presentation will closely follow that of \cite{Nicolai:2001ac}.  Apart from the graviton, this theory has eight gravitini, $\psi_\mu^A$, the gauge connections, $B_\mu{}^\cM$, $8n$ fermions, $\chi^{\dot A r}$, and $8n$ bosons.  The supersymmetries, $\epsilon^A$, and the gravitini transform in the, $8^+$, representation of $SO(8)$ and the fermions, $\chi^{\dot A r}$, transform in the opposite helcity, $8^-$, representation of $SO(8)$.

As with most gauged supergravity theories,  the starting point is the scalar manifold and its coupling to gauge fields.   The scalar coset is
\begin{equation}
\frac{G}{H} ~\equiv~ \frac{SO(8,n)}{SO(8) \times SO(n)} \,.
\label{coset1}
\end{equation}
The scalar fields are then parametrized by a coset representative, $L(x)$, that is viewed as transforming on the left under a global action of $g \in G$, and on the right under a composite, local symmetry $h(x) \in H$:  $L(x) \to g L(x) h (x)^{-1}$.  The gauge group, $G_0$, which we will specify later, is a subgroup of $G$ and the gauging promotes  $G_0$ to a local  action on $L(x)$: 
\begin{equation}
L(x)\; \longrightarrow \; g_0 (x)\, L(x) \,h^{-1}(x)\;, \qquad
  g_0 (x) \in G_0 \; , \; h(x) \in H \;,
\label{GHaction}
\end{equation}

The gauge subgroup and its coupling are defined by an embedding tensor, $\Theta_{\cM\cN}$:
\begin{equation}
\widehat \cD_\mu ~\equiv~ \partial_\mu~+~ g\, \Theta_{\cM\cN}\,B_\mu{}^\cM t^\cN  \,,
\label{covDer0}
\end{equation}
where $B_\mu{}^\cM $ are the gauge connections.

The  covariant derivative of $L(x)$  is then used to define various connection components via a Lie algebra decomposition\footnote{This is where the signs of the generators are critical.}:
\begin{equation}
L^{-1}\left(\partial_\mu~+~ g\, \Theta_{\cM\cN}\,B_\mu{}^\cM t^\cN\right) L ~\equiv~  \coeff{1}{2}\, \cQ_\mu^{IJ} \,X^{IJ}  + \coeff{1}{2}\, \cQ_\mu^{rs} X^{rs} +  \cP_\mu^{Ir} \,X^{Ir}
\;,
\label{eq:LinvDL}
\end{equation}
The tensor $\cP_\mu^{Ir}$ will define the bosonic kinetic term and  the $\cQ_\mu$'s are used to define covariant derivatives on fermions:
\begin{equation}
\begin{aligned}
D_\mu \, \psi_\nu^A ~\equiv~ &   \partial_\mu \psi_\nu^A ~-~  \widehat \Gamma^\rho_{\mu \nu}\, \psi_\rho^A ~+~ \coeff{1}{4}\,\omega_\mu{}^{ab}\,\gamma_{ab}\, \psi_\nu^A  + \coeff{1}{4}\,\cQ_\mu^{IJ}\Gamma^{IJ}_{AB} \psi_\nu^B \,,   \\
D_\mu \chi^{\dot{A} r} ~\equiv~ &  \partial_\mu \chi^{\dot{A} r} ~+~  \coeff{1}{4}\,\omega_\mu{}^{ab}\,\gamma_{ab}\, \chi^{\dot{A} r}  ~+~
\coeff{1}{4}\, \cQ_\mu^{IJ} \Gamma^{IJ}_{\dot{A}\dot{B}} \,\chi^{\dot{B} r} ~+~  \cQ_\mu^{rs} \,\chi^{\dot{A} s} \;.
\end{aligned}
\nonumber
\end{equation}
Here $\widehat \Gamma^\rho_{\mu \nu}$ is the Christofflel connection and should not be confused with the $\Gamma$-matrices.  This term is omitted in \cite{Nicolai:2001ac}, but this omission is harmless because only the skew derivative $D_{[ \mu} \, \psi_{\nu ]}^A$ appears in the action and so the Christoffel connection disappears.

The scalar fields enter the action and supersymmetry variations in several non-trivial ways, and these are characterized by various $A$-tensors that are derived from the $T$-tensor.  To define the latter one needs to decompose the adjoint action of the scalar matrix:
\begin{equation}
L^{-1} t^\cM L ~\equiv~   {\cV^{\cM}}_{\cA}  \,t^\cA ~=~  \coeff{1}{2}\,{\cV^{\cM}}{}_{IJ}   \,X^{IJ}  ~+~  \coeff{1}{2}\, {\cV^{\cM}}{}_{rs}  \,X^{rs} ~+~  {\cV^{\cM}}{}_{Ir} \,X^{Ir} \,, 
\label{adjaction}
\end{equation}
and then the  $T$-tensor is defined by: 
\begin{equation}
T_ {\cA | \cB} ~\equiv~ \Theta_{\cM \cN}\, {\cV^{\cM}}_{\cA}  \, {\cV^{\cN}}_{\cB}  \,.
\label{Ttensor}
\end{equation}
The $A$-tensors are then constructed from various pieces of the $T$-tensor:
\begin{equation}
\begin{aligned}
A_1^{AB} ~=~& -\delta^{AB}\,\theta -  \coeff{1}{48}\,\Gamma^{IJKL}_{AB}\,T_ {IJ | KL}  \,,  \\
A_2^{A\dot{A} r} ~=~& -\coeff{1}{12}\,\Gamma^{IJK}_{A\dot{A}}\,T_ {IJ | Kr}  \,, \\ 
A_3^{\dot{A} r\, \dot{B} s} ~=~& 2\delta^{\dot{A}\dot{B}}\delta^{rs}\,\theta + \coeff{1}{48}\,\delta^{rs}\,\Gamma^{IJKL}_{\dot{A}\dot{B}}\,T_ {IJ | KL}
+\coeff{1}{2}\,\Gamma^{IJ}_{\dot{A}\dot{B}}\,T_ {IJ | rs} \,.
\end{aligned}
\label{Atensors}
\end{equation}
where  $\theta  \equiv  \frac{2}{(8+ n)(7+ n)}\, \eta^{\cM\cN}\Theta_{\cM\cN}$, and $\eta^{\cM\cN}$ is the Cartan-Killing form on $G$.  For more details, see \cite{Nicolai:2001ac}.  In the gauging we consider here, one has $\theta =0$.

The Lagrangian is then given by:

\begin{equation}
\begin{aligned}
\cL ~=~ & -\coeff{1}{4} \,e\,R
~+~ \coeff{1}{2}  \,\epsilon^{\mu\nu\rho} \, \overline{\psi}^A_\mu D_\nu\psi^A_\rho
~+~ \coeff{1}{4}\,e  \,\cP_\mu^{Ir}  \,\cP^\mu{}^{\,Ir} 
~-~  \coeff{1}{2}\,i e \, \overline{\chi}^{\dot{A} r}\gamma^\mu D_\mu\chi^{\dot{A} r} 
\nonumber\\[1ex]
&~-~  \coeff{1}{4}\,\epsilon^{\mu\nu\rho}\,g\, \Theta_{\cM\cN}\,B_\mu{}^\cM
\Big(\partial_\nu B_\rho\,{}^\cN
~+~ \coeff{1}{3}\,g \, \Theta_{\cK\cL}\, f^{\cN\cK}{}_{\cP}\,
B_\nu{}^\cL B_\rho{}^\cP \Big) 
\nonumber\\  
&~-~ 
\coeff{1}{2}\,e\, \cP_\mu^{Ir} \, \overline{\chi}^{\dot{A} r} \, \Gamma^I_{A\dot{A}}\gamma^\nu\gamma^\mu\psi^A_\nu
~+~ \coeff{1}{2}\,g\, e \, A_1^{AB} \,\overline{\psi}{}^A_\mu \gamma^{\mu\nu} \psi^B_\nu \nonumber\\
&  
~+~  i\,g\,e \,A_2^{A\dot{A} r} \,\overline{\chi}^{\dot{A} r} \gamma^{\mu} \psi^A_\mu
~+~ \coeff{1}{2}\,g\,e \,A_3^{\dot{A} r\, \dot{B} s}\, \overline{\chi}^{\dot{A} r}\chi^{\dot{B} s}  ~-~e\, V \,,
\end{aligned}
\label{eq:3Daction}
\end{equation}
where the potential is defined by:
\begin{equation}
V~=~ - \frac{1}{4}\,g^2 \, \Big(A_1^{AB}A_1^{AB}-\coeff{1}{2}\,A_2^{A\dot{A} r}A_2^{A\dot{A} r}\Big) \,.
\label{potential1}
\end{equation}
One should note that we have replaced the potential $W$ of \cite{Nicolai:2001ac} by $-V$ as we wish to avoid confusion with the superpotential that will be defined later.

This action is then invariant under the supersymmetry transformations  \cite{Nicolai:2001ac}:
\begin{align}
L^{-1}\delta L ~=~ & X^{Ir}\, \overline{\epsilon}^A\,\Gamma^I_{A\dot{A}}\, \chi^{\dot{A} r} \,,  \qquad 
\delta\chi^{\dot{A} r} ~=~  \coeff{1}{2}\,i \,\Gamma^I_{A\dot{A}}\gamma^\mu\epsilon^A \, \cP_\mu^{Ir}  ~+~ g \, A_2^{A\dot{A} r}\epsilon^A \,, \label{susy1} \\[1ex]
\delta e_\mu{}^\alpha  ~=~ &  i \, \overline{\epsilon}^A\,\gamma^\alpha\psi^A_\mu \,, \qquad   \ \  \qquad \delta\psi^A_\mu ~=~  D_\mu \epsilon^A ~+~ i \,g \,A_1^{AB}\gamma_\mu\epsilon^B \,,  \label{susy2} \\[1ex]
\delta B_\mu{}^\cM   ~=~ &  -\coeff{1}{2} \,{\cV^{\cM}}{}_{IJ}\,\overline{\epsilon}^A\,\Gamma^{IJ}_{AB}\, \psi^B_\mu ~+~  i\,{\cV^{\cM}}{}_{Ir}\,\overline{\epsilon}^A\,\Gamma^I_{A\dot{A}}\gamma_\mu \, \chi^{\dot{A} r} \,. \label{susy3}
\end{align}

As usual there are also four-fermion terms in the action and higher fermion terms in the supersymmetry variations. According to the arguments in  \cite{Nicolai:2001ac}, these terms are the same as those given in \cite{Marcus:1983hb}.  Since we are interested in the BPS equations of supersymmetric bosonic backgrounds, we will not need any of these higher fermion terms.

\subsection{Truncating to the  $\Neql{4}$ gauged supergravity}
\label{sec:Neql4truncation}

The $\Neql{4}$ theory has half the supersymmetries and, at a minimum, this means removing half the gravitini and their superpartners.  In six dimensions, the superpartners of the truncated gravitini are  four of the self-dual tensors, whose removal translates, in three dimensions, to reducing the $SO(8,n)$ global symmetry of the ungauged theory to $SO(4,n)$.  This reduction also reduces the number of scalars from $8n$ to $4n$, which means halving the number of fermions, $\chi^{\dot A r}$.   The coset (\ref{coset1}) is thus reduced to:
\begin{equation}
\frac{\hat G}{\hat H} ~\equiv~ \frac{SO(4,n)}{SO(4) \times SO(n)} \,.
\label{coset2}
\end{equation}

The $SO(4,n)$  embeds in the obvious manner into  $SO(8,n)$ and indeed, in this paper we will simply view the $\Neql{4}$ theory as embedded in the larger $\Neql{8}$ theory.  We will therefore consider the scalar matrix to be that of the $SO(8,n)$ theory but non-trivial only in the $SO(4,n)$ block defined by $I,J, \ldots  =1,2,3,4$.  

The corresponding truncation of the fermions is easily implemented: One must require 
\begin{equation}
\big(\oneone  ~-~ \Gamma^{5678}\big) \,  \Phi ~=~ 0\,,
\label{SO8proj1}
\end{equation}
where $\Phi$ is any fermion, including the supersymmetries.  Because the $SO(8)$ helicity projector is $\Gamma^{12345678}$, this condition translates to:
\begin{equation}
\big(\oneone  ~-~ \Gamma^{1234}\big)_{AB} \, \epsilon^B  ~=~ 0 \,, \qquad \big(\oneone  ~-~ \Gamma^{1234}\big)_{AB} \,\psi^B_\mu  ~=~ 0 \,, \qquad \big(\oneone  ~+~ \Gamma^{1234}\big)_{\dot A \dot B}\, \chi^{\dot B r}  ~=~ 0   \,.
\label{SO8proj2}
\end{equation}
This cuts the supersymmetries, and all the fermionic degrees of freedom, in half.

It is easy to see that this truncation is consistent with the supersymmetry variations. Because of the restriction on the scalars, the Lie algebra matrices in (\ref{adjaction}) must live in $SO(4,n)$.  This means that all the indices on the $T$-tensor and on  $\cQ_\mu^{IJ}$ and $\cP_\mu^{Ir}$ must lie in  the Lie algebra of $SO(4,n)$.  It follows that  $\cQ_\mu^{IJ} \Gamma^{IJ}$, $\cP_\mu^{Ir} \Gamma^{I}$,  and all the $A$-tensors  commute with $\Gamma^{5678}$.  This implies that the supersymmetry variations of the fermions respect the projection (\ref{SO8proj1}).  The  variation $L^{-1}\delta L$ is easily seen to vanish along $X^{Ir}$, $I=5, \dots, 8$ as a consequence of  (\ref{SO8proj1}).  Finally, the fact that $\cV^{\cM}{}_\cA$ lies in $SO(4,n)$ means that it is consistent with the supersymmery variations to restrict the gauge fields, $B_\mu{}^\cM$,  to a sub-algebra of $SO(4,n)$.

There is a simple way to characterize this truncation in terms of a group invariant sector of the $\Neql8$ theory.  There is an $(SU(2))^4$ subgroup of $SO(8)$.  One of these $SU(2)$ rotations is characterized as the self-dual rotations on the indices $5,6,7,8$.  The projection condition (\ref{SO8proj1}) requires all the fermions to be singlets under this $SU(2)$. Moreover, the centralizer of this $SU(2)$ in $SO(8,n)$ is precisely $SO(4,n)$.   Thus the truncation of the $\Neql 8$ theory to the $\Neql4$ theory may be defined as reducing to the singlet sector of this $SU(2)$.

While the $\Neql4$  theory was the starting point of  \cite{Mayerson:2020tcl}, and can be directly related to superstrata, nothing prevents us from considering the three-dimensional solutions presented in \cite{Mayerson:2020tcl} as being part of the larger $\Neql8$ theory and analyzing the supersymmetry from that perspective.  We will take this approach and find that the projection condition  (\ref{SO8proj2})
emerges from the analysis of the supersymmetry of the solution.  

\subsection{The minimal couplings of  $\Neql{4}$ gauged supergravity}
\label{sec:Neql4couplings}

The three-dimensional theory that underlies the superstrata is the $SO(4,5)$ theory described in  \cite{Mayerson:2020tcl}.  Here we will summarize the essential features of that theory.  Our presentation will differ slightly from   \cite{Mayerson:2020tcl} because of our conventions and because  we will fix some of the parameters\footnote{This is not a restriction because these parameters were ultimately fixed in  \cite{Mayerson:2020tcl} by requiring that the superstratum data solved the equations of motion.} that appeared in   \cite{Mayerson:2020tcl}.

First, it is most convenient to express the embedding tensor details in terms of the $GL(4\,, \IR)$ basis of $SO(4,5)$ in which the invariant matrix takes the form: 
\begin{equation}
\hat \eta ~\equiv~
\left( \begin{matrix} 
0_{4 \times 4} & \oneone_{4 \times 4} & 0 \\
\oneone_{4 \times 4} & 0_{4 \times 4} & 0 \\
0 &0& -1
\end{matrix} \right) \,.
\label{invmat2}
\end{equation}
Note that we have fixed the parameter $\varepsilon$ of  \cite{Mayerson:2020tcl} by taking:
\begin{equation}
\varepsilon ~=~-1  \,.
\label{varepschoice}
\end{equation}
This is merely a  choice of convention.   The change of basis matrices to go back to canonical  
$SO(4,5)$ conventions with $\eta$ of the form (\ref{invmat1}) may be found in the Appendix.

Following \cite{Mayerson:2020tcl}, a vector of $SO(4,5)$ will be denoted by 
\begin{equation}
\cX_{\bar M}  ~\equiv~ ( \cX_I, \cX^I, \cX_0) \,,   \qquad \cX^{\bar M}  ~\equiv~ ( \cX^I, \cX _I, \varepsilon \cX_0) \,,
\label{Ghatvec}
\end{equation}
where the indices are raised and lowered using (\ref{invmat2}).  The components, $\cX_I$ and $\cX^I$, transform, respectively, in the $4$ and $\overline 4$ of $GL(4,\IR)$.  

The embedding tensor, $\Theta$, is totally anti-symmetric\footnote{This is why $\theta$ in (\ref{Atensors}) vanishes.}:   
\begin{equation}
\Theta_{\bar K \bar L, \bar M \bar N  } ~=~ \theta_{\bar K \bar L  \bar M \bar N }  ~=~  \theta_{[\bar K \bar L  \bar M \bar N ]}  \,,
\end{equation}
and the only non-vanishing pieces are \cite{Nicolai:2003ux,Deger:2014ofa,Samtleben:2019zrh,Mayerson:2020tcl}:   
\begin{equation}
 \theta_{IJKL }  ~=~ - g_0 \,   \epsilon_{IJKL }  \,, \qquad  {\theta_{IJK }}^L  ~=~  \coeff{1}{2}\, g_0 \,    \epsilon_{IJKM} \, \delta^{LM}   \,,
\label{EmbTens}
\end{equation}
One should note that we have replaced the parameters  $\alpha$  and $\gamma_0$ in   \cite{Mayerson:2020tcl} according to:
\begin{equation}
\alpha ~=~ \gamma_0 ~=~     \coeff{1}{2}\, g_0
\label{gammalphrepl}
\end{equation}
These replacements follow from equations (3.7)  and (4.5)  of \cite{Mayerson:2020tcl} with $\varepsilon =-1$, as  in (\ref{varepschoice}).
We also note the $g_0$ is related to the supergravity charges of the D1-D5 compactification via:
 \begin{equation}
g_0 ~\equiv~ (Q_1 Q_5)^{-\frac{1}{4}} \,.
\end{equation}
(See equation (4.5) of  \cite{Mayerson:2020tcl}.)

The gauge group is $SO(4) \ltimes \IT^6$  and the connection ${A_\mu}^{\bar K \bar L}$ has the following $12$ fields:
\begin{equation}
 {A_\mu}^{IJ} ~=~ - {A_\mu}^{JI}   \,, \qquad  A_{\mu}{}^{J}{}_I ~=~  -{A_{\mu \,J}}^I \,.
\end{equation}
Because of the $\epsilon$-symbols in the embedding tensor, it is convenient to define:
\begin{equation}
{\widetilde A_\mu}{}^{IJ}  ~\equiv~ \coeff{1}{2} \,\epsilon_{IJKL}\,{A_\mu}^{KL} \,, \qquad \ {\widehat A_\mu}{}^{IJ}  ~\equiv~ \coeff{1}{2} \,\epsilon_{IJKL}\,{{A_\mu}{}^K}{}_L   \,, 
\label{dualGFs}
\end{equation}
As in \cite{Mayerson:2020tcl}, we  introduce:
\begin{equation}
{B_\mu}{}^{IJ}  ~\equiv~ 4\,g_0\, \big(\, {\widetilde A_\mu}{}^{IJ}    -  {\widehat A_\mu}{}^{IJ} \big)  \,.
\label{Bvecden}
\end{equation}

The gauge connection acts according to (\ref{covDer0}), which we now write as:
\begin{equation}
\begin{aligned}
\widehat\cD_\mu \, \cX_{\bar P}  ~\equiv~&  \partial_\mu \, \cX_{\bar P}  ~+~ g\,  {A_\mu}^{\bar K \bar L} \, \Theta_{\bar K \bar L, \bar M \bar N } \, \big( \,X^{\bar M \bar N} \, \big)_{\bar P}{}^{\bar Q} (\cX_{\bar Q} ) \\
  ~=~ & \partial_\mu \, \cX_{\bar P}  ~+~g\,  {A_\mu}^{\bar K \bar L} \,\big( \Theta_{\bar K \bar L, \bar P \bar N } \, \hat \eta^{\bar N\bar Q} \, \cX_{\bar Q} -  \Theta_{\bar K \bar L, \bar M \bar P } \, \hat \eta^{\bar M \bar Q} \, \cX_{\bar Q} \big )  
 \,, 
 \end{aligned}
\label{covderiv1}
\end{equation}
where we have used  (\ref{matform}).  There are several things to note at this point.  Because we have replaced adjoint indices by doubled indices like, $\bar K \bar L$, we are double summing over the adjoint representation.  One can take this to be part of the definition and normalization of the components of the embedding tensor, $\Theta_{\bar K \bar L, \bar M \bar N } $.  Indeed one can compensate for the double sums by sending  $\Theta \to \frac{1}{4} \Theta$.  Next, compared to  \cite{Mayerson:2020tcl}, we have introduced another gauge coupling, $g$, (inherited from (\ref{covDer0})) and we   are using the opposite signs for the group generators and structure constants.  In principle, this will change the signs of the gauge couplings throughout the action.  However we are now going to choose 
\begin{equation}
g  ~=~    -1 \,.
\label{gchoice}
\end{equation}
This will compensate for all the opposite signs compared to  \cite{Mayerson:2020tcl}, and lead to precisely the same covariant derivatives and actions.  Indeed, exactly as in \cite{Mayerson:2020tcl}, our covariant derivative on a $SO(4,5)$ vector, (\ref{Ghatvec}), in the $GL(4, \IR)$ basis becomes
\begin{equation}
\begin{aligned}
\widehat\cD_\mu \, \cX_{I}  &~=~ \partial_\mu \, \cX_{I}  ~+~ {B_\mu}{}^{IJ}  \,  \cX^{J}  ~-~  2\, g_0 \,\widetilde A_\mu{}^{IJ} \, \cX_{J} \,, \\  
\widehat\cD_\mu \, \cX^{I}  &~=~ \partial_\mu \, \cX^{I}    ~-~  2\, g_0 \,\widetilde A_\mu{}^{IJ} \, \cX^{J} \,, \qquad \widehat\cD_\mu \, \cX_0 ~=~ \partial_\mu \, \cX_0  \,.
\end{aligned}
\label{covderiv2}
\end{equation}

It may seem circuitous to have used the opposite-sign generators for $SO(4,5)$ only to undo this choice through (\ref{gchoice}).  However, the signs of the generators are also crucial to the definition of all the tensors in Section \ref{sec:Neql8} and so we have taken this apparently circuitous route so as to arrive at the action of \cite{Mayerson:2020tcl} while respecting the conventions essential to \cite{Nicolai:2001ac}.

The connections, ${B_\mu}{}^{IJ}$, lie in the upper triangular part of the $SO(4,5)$ matrices:
\begin{equation}
\cB_\mu ~\equiv~
\left( \begin{matrix} 
0 & {B_\mu}{}^{IJ} & 0\\
0 & 0 & 0 \\
0 &0 & 0
\end{matrix} \right) \,.
\end{equation}
These are  the gauge fields of $\IT^6$ and will ultimately be integrated out of the action.  The vector fields $A_\mu{}^{IJ}$ are those of $SO(4)$ but they act with their duals, and with a gauge coupling of $-2 g_0$.  It is therefore useful to introduce the $SO(4)$ covariant derivatives:
\begin{equation}
\begin{aligned}
\cD_\mu \, \cX_{I}  &~=~ \partial_\mu \, \cX_{I}   ~-~  2\, g_0\,\widetilde A_\mu{}^{IJ} \, \cX_{J} \,, \\  
 \cD_\mu \, \cX^{I}  &~=~ \partial_\mu \, \cX^{I}    ~-~  2\, g_0\,\widetilde A_\mu{}^{IJ} \, \cX^{J} \,, \qquad  \cD_\mu \, \cX_0 ~=~ \partial_\mu \, \cX_0  \,.
\end{aligned}
\label{covderiv4}
\end{equation}
%

\subsection{The Maxwell fields }
\label{sec:Maxwell}

The connections, (\ref{covderiv4}), lead to the Maxwell fields 
\begin{equation}
F_{\mu \nu}{}^{IJ}  ~=~ \coeff{1}{2}\, \epsilon_{IJKL} \,  \widetilde F_{\mu \nu}{}^{KL}    ~=~  \partial_{\mu}   A_{\nu}{}^{IJ}   ~-~  \partial_{\nu}   A_{\mu}{}^{IJ}   ~-~ 2 \,  g_0 \, \big(   A_{\mu} {}^{IL} \,\widetilde  A_{\nu} {}^{LJ}  ~-~ A_{\mu} {}^{JL} \,\widetilde  A_{\nu} {}^{LI}\big)  \,.
\label{fieldstrength2}
\end{equation}
The Chern-Simons action appearing in  (\ref{eq:3Daction}) is:
\begin{equation}
\begin{aligned}
& - \coeff{1}{4}\,\epsilon^{\mu\nu\rho}\,g\, \Theta_{\cM\cN}\,B_\mu{}^\cM \Big(\partial_\nu B_\rho\,{}^\cN ~+~ \coeff{1}{3}\,g \, \Theta_{\cK\cL}\, f^{\cN\cK}{}_{\cP}\, B_\nu{}^\cL B_\rho{}^\cP \Big) \\
 &~=~  + \coeff{1}{4}\,\epsilon^{\mu\nu\rho}\,\Theta_{\cM\cN}\,B_\mu{}^\cM \Big(\partial_\nu B_\rho\,{}^\cN ~-~ \coeff{1}{3} \, \Theta_{\cK\cL}\, f^{\cN\cK}{}_{\cP}\, B_\nu{}^\cL B_\rho{}^\cP \Big)
 \,,
\end{aligned}
\label{CSact1}
\end{equation}
where we have set $g=-1$.  Using the double sum conventions, this translates   into:
\begin{equation}
\cL_{CS} ~=~  \coeff{1}{4}\, \varepsilon^{\mu \nu \rho} \,  A_\mu{}^{\bar K\bar L} \,\Theta_{\bar K\bar L,\bar M\bar N}\,
\Big( \partial_\nu A_\rho {}^{\bar M\bar N}  ~-~  \coeff{1}{3}\, f^{\bar M\bar N,\bar P\bar Q}{}_{\bar R\bar S}\,\Theta_{\bar P\bar Q,\bar U\bar V}\,A_\nu{}^{\bar U\bar V} A_\rho{}^{\bar R\bar S}\Big)\,.
\label{CSgeneric}
\end{equation}
To compensate for the double sums one can  rescale $\Theta \to \frac{1}{4} \Theta$, as one does in going from (\ref{covDer0}) to (\ref{covderiv1}).   This leads to the correct normalization in the first term, however this introduces a factor of $\frac{1}{16}$ in the second term whereas there are five double sums. The extra factor of $\frac{1}{2}$ is, however, built in through our definition of the structure constants in (\ref{comms1})  and the resulting factors of $\frac{1}{2}$ in (\ref{structureconsts}).  One should also note that the second term in (\ref{CSgeneric}) has the  opposite sign to that of \cite{Mayerson:2020tcl}.  This is because our structure constants also have the opposite sign.

Using   (\ref{structureconsts}),  (\ref{EmbTens}) and   (\ref{Bvecden})  we arrive at
\begin{equation}
\cL_{CS}  ~=~  \coeff{1}{2}\, \varepsilon^{\mu \nu \rho} \, \Big[  g_0 \,\big(A_\mu{}^{IJ}\, \partial_\nu  \widetilde A_\rho{}^{IJ}  ~+~\coeff{4}{3}\,  g_0 \, A_\mu{}^{IJ} \,  A_\nu{}^{JK}\, A_\rho{}^{KI} \,\big) ~-~  \coeff{1}{4}\,  {B_\mu}{}^{IJ}  \, F_{\nu \rho}^{IJ} \Big] \,,
\label{CSterm}
\end{equation}
which exactly matches\footnote{We have re-ordered some of the indices relative to  the expression in \cite{Mayerson:2020tcl}.} (2.33) of \cite{Mayerson:2020tcl}.

\subsection{The scalar fields }
\label{sec:Scalars}

Following \cite{Mayerson:2020tcl}, our scalar matrix will be defined by:
\begin{equation}
\begin{aligned}
{{L}_{\bar M}}^{\bar K} & ~=~
\left( \begin{matrix} 
{P_I}^J &  \frac{1}{2} \, \chi_I \,  \big({(P^{-1})_J}^K \chi_K\big) & \chi_I  \\
0& {(P^{-1})_J}^I & 0 \\
0 &{(P^{-1})_J}^K \chi_K & 1
\end{matrix} \right)  \,,
\end{aligned}
\label{scalmat}
\end{equation}
where $P =P^T$ is a symmetric $GL(4,\IR)$ matrix. The matrix, $P$, can be chosen to be symmetric because of the composite local symmetry, $H$, in (\ref{GHaction}).  Since we also have a local $SO(4)$ gauge symmetry, it  will be convenient  to diagonalize $P$ in terms of four scalar fields, $\mu_i$:
\begin{equation}
P  ~=~  {\rm diag} \big(\,  e^{\mu_1} \,, \,  e^{\mu_2} \,, \,  e^{\mu_3} \,, \,  e^{\mu_4} \, \big) \,.
\label{Pdiag}
\end{equation}
One should remember that the  gauge symmetry only acts on the left of $L$ in (\ref{GHaction}), and this translates into the purely left action of the gauge fields in (\ref{eq:LinvDL}), which, in turn, means that the covariant derivative, $\cD_\mu$, only acts on the left of $P$.

It is also convenient to define the scalar matrix $m_{IJ}$  and its inverse,  $m^{IJ}$:
\begin{equation}
m_{IJ}   ~\equiv~   \big(P  \, P^T\big)_{IJ} \,, \qquad m^{IJ}   ~=~  \big ( (P^{-1})^T\,P^{-1}  \big)^{IJ}  \,.
\label{Mdefn}
\end{equation}
One should note that the covariant derivative of $m$ is therefore given by:
\begin{equation}
\cD_\mu m_{IJ}   ~=~   \partial_\mu m_{IJ}  ~-~   2\, g_0\,\widetilde A_\mu{}^{IK} m_{KJ }~-~   2\, g_0\,\widetilde A_\mu{}^{JK} m_{IK }   \,   \,.
\label{Dmform}
\end{equation}

We will also define the following combinations of fields:
\begin{equation}
\begin{aligned}
Y_{\mu \, IJ}  ~\equiv~&  \chi_J \,\cD_\mu \chi_I ~-~  \chi_I \,\cD_\mu\chi_J \,, \\
C_{\mu}^{IJ}  ~\equiv~  & {B_\mu}{}^{IJ} ~+~  \coeff{1}{2}\, Y_{\mu \, IJ}  \,, \qquad  \qquad  \scrC_{\mu}^{IJ}   ~\equiv~  P^{-1}{}_{I}{}^K \,  P^{-1}{}_{J}{}^L \, C_\mu^{KL}\,.
\end{aligned}
\label{cs_definition}
\end{equation}
Note that these objects break the $GL(4\,, \IR)$ covariance and are to be considered only as $SO(4)$ tensors.  This means that we will not distinguish raised and lowered indices for such objects.

The various pieces of the Lie algebra element (\ref{eq:LinvDL}) are then given by:
\begin{equation}
\begin{aligned}
\cQ_\mu^{IJ}   ~=~ & \coeff{1}{2}\, \Big[ \, \big(P^{-1} \cD_\mu \, P\big)_{I}{ }^{\, J}  ~-~  \big(P^{-1} \cD_\mu \, P\big)_{J}{}^{\, I} ~+~\scrC_{\mu}^{IJ} \   \,  \Big] \,, \\
\cP_\mu^{Ir}   ~=~ & \coeff{1}{2}\, \Big[ \, \big(P^{-1} \cD_\mu \, P\big)_{I}{ }^{\, r}  ~+~  \big(P^{-1} \cD_\mu \, P\big)_{r}{}^{\, I} ~-~\scrC_{\mu}^{Ir} \   \,  \Big] \,, \qquad 1 \le r  \le 4 \,, \\
\cQ_\mu^{rs}   ~=~ & - \coeff{1}{2}\, \Big[ \, \big(P^{-1} \cD_\mu \, P\big)_{r}{ }^{\, s}  ~-~  \big(P^{-1} \cD_\mu \, P\big)_{s}{}^{\, r} ~+~\scrC_{\mu}^{rs} \   \,  \Big] \,,  \qquad 1 \le r,s  \le 4 \,,\\
\cQ_\mu^{r5}   ~=~&   \coeff{1}{\sqrt{2}}\, \big(P^{-1} \big)_{r}{ }^{\, J}  \, \cD_\mu \chi_J  \,,  \qquad
\cP_\mu^{I5}   ~=~  - \coeff{1}{\sqrt{2}}\, \big(P^{-1} \big)_{I}{ }^{\, J}  \,\cD_\mu  \chi_J \,,  \qquad 1 \le r  \le 4  \,.
 \end{aligned}
\label{ScalarDers}
\end{equation}
Note that in defining $\cQ_\mu^{rs}$ we are using the generators, $X^{rs}$,  defined by (\ref{comms1}) or (\ref{matform}), and not those of \cite{Nicolai:2003ux}.

To define the $A$-tensors in the supersymmetry variations, we introduce the superpotential:
\begin{equation}
\begin{aligned}
W  ~\equiv~ & \coeff{1}{4} \, g_0 \,  (\det(P))^{-1}  \,   \Big [\, 2 \,\Big(1- \coeff{1}{4} \,  (\chi_A \chi_A)\Big) ~-~ {\rm Tr}\big(P\, P^T\big)  \, \Big] \\
~=~ & \coeff{1}{4} \, g_0 \,e^{-\mu_1 -\mu_2-\mu_3-\mu_4} \, \Big [\, 2 \,\Big(1- \coeff{1}{4} \,  (\chi_A \chi_A)\Big) ~-~ \Big( e^{2\, \mu_1}+e^{2\, \mu_2}+e^{2\, \mu_3}+e^{2\, \mu_4}  \Big) \, \Big]   \,,
 \end{aligned}
\label{superpot}
\end{equation}
Then we find:
\begin{equation}
\begin{aligned}
A_1^{AB}   ~=~ W\, (\Gamma^{1234}\big)_{AB} \,, \qquad A_2^{A\dot A \, r}   ~=~ & \frac{\partial W}{\partial \mu_r} \, (\Gamma^{1234}\,\Gamma^{r}\big)_{A \dot A} \,, \quad 1 \le r \le 4  \,, \\
\qquad A_2^{A\dot A \, 5}   ~=~ &- \sqrt{2}\, \sum_{j=1}^4 \,e^{\mu_j} \, \frac{\partial W}{\partial \chi_j} \, (\Gamma^{1234}\,\Gamma^{j}\big)_{A \dot A}  \,,
\end{aligned}
\label{A1A2tens}
\end{equation}
where there is no sum on $r$ in the expression for $A_2^{A\dot A \, r}$.  While we will not need it, we also find:
\begin{equation}
\begin{aligned}
A_3^{\dot A  r\, \dot B s}   ~=~ & - W\, \delta^{rs}\, (\Gamma^{1234}\big)_{\dot A \dot B}  ~-~ \epsilon_{IJrs}  \,\frac{\partial^2 W}{\partial \mu_r \, \partial \mu_s}   (\Gamma^{IJ}\big)_{\dot A \dot B}  \,, \qquad 1 \le r, s \le 4  \,,\\
A_3^{\dot A  r\, \dot B 5}   ~=~ &  \sqrt{2}\,  \sum_{j=1}^4 \,e^{\mu_j} \,(\Gamma^{IJ}\big)_{\dot A \dot B}   \, \epsilon_{IJ r j}  \,  e^{\mu_j}  \,\frac{\partial^2 W}{\partial \mu_r \, \partial \chi_j}  \,, \qquad 1 \le r  \le 4  \,, \\
 A_3^{\dot A  5\, \dot B 5}   ~=~&  - W\,(\Gamma^{1234}\big)_{\dot A \dot B}  \,.
\end{aligned}
\label{A3tens}
\end{equation}
where there is no sum on $r$ or $s$.

Using these expressions we find that the potential (\ref{potential1}) is given by:
\begin{equation}
V~=~ \delta^{ij} \frac{\partial W}{\partial \mu_i}  \frac{\partial W}{\partial \mu_j}  ~+~ 2\, m^{IJ} \, \frac{\partial W}{\partial \chi_I} \frac{\partial W}{\partial \chi_J}   ~-~2\, W^2  \,.
\label{potential2}
\end{equation}
Thus the scalar sector of the theory, and its role in the supersymmetries, are  determined entirely by the superpotential (\ref{superpot}).  

Using the explicit forms of $P$, or $m$, we find 
\begin{equation}
\begin{aligned}
V~=~ & \coeff{1}{4}\, g_0^2   \,e^{-2\, (\mu_1 +\mu_2+\mu_3+\mu_4)} \, \Big [\, 2 \,\big(1- \coeff{1}{4} \,  (\chi_I \chi_I)\big)^2    ~+~ \big( e^{4\, \mu_1}+e^{4\, \mu_2}+e^{4\, \mu_3}+e^{4\, \mu_4}  \big)  \\
& \qquad\qquad\qquad\qquad \qquad\qquad~+~\coeff{1}{2}\, \big( e^{2\, \mu_1}\, \chi_1^2 +e^{2\, \mu_2}\, \chi_2^2 + e^{2\, \mu_3}\, \chi_3^2+e^{2\, \mu_4}\, \chi_4^2  \big) \\
& \qquad\qquad\qquad\qquad \qquad\qquad~-~\coeff{1}{2}\, \big( e^{2\, \mu_1} +e^{2\, \mu_2} + e^{2\, \mu_3} + e^{2\, \mu_4}  \big)^2  \, \Big]  \\
~=~ & \coeff{1}{4}\, g_0^2   \,  \det\big(m^{IJ}\big) \, \Big [\, 2 \,\big(1- \coeff{1}{4} \,  (\chi_I \chi_I)\big)^2    ~+~ m_{IJ} m_{IJ}  ~+~\coeff{1}{2} \,  m_{IJ} \chi_I \chi_J  ~-~\coeff{1}{2} \,  m_{II}  \,  m_{JJ}\, \Big]
 \,.
\end{aligned}
\label{potential3}
\end{equation}
Using (\ref{Pdiag}) and (\ref{ScalarDers}),  the scalar kinetic term can be written
\begin{equation}
\begin{aligned}
\cP_\mu^{Ir}  \,\cP^\mu{}^{\,Ir}  ~=~   g^{\mu \nu} \,\Big[ & \,  \coeff{1}{4} \, \big( m^{IK} \, \cD_\mu\, m_{KJ}  \big)   \big( m^{JL} \, \cD_\nu\, m_{LI}  \big) \\
  &  ~+~\coeff{1}{2} \, m^{IJ}   \, (\cD_\mu\, \chi_{I})  \, (\cD_\nu\, \chi_{J})   ~+~  \coeff{1}{4} \, \big( m^{IJ} \, m^{KL} \, C_{\mu}^{IK}  \, C_{\nu}^{JL}   \big) \, \Big] \,.
\end{aligned}
\label{scalkin}
\end{equation}
Thus the expressions (\ref{potential3}) and (\ref{scalkin}) precisely match the corresponding quantities in \cite{Samtleben:2019zrh,Mayerson:2020tcl}.

\subsection{The three-dimensional supergravity action}
\label{sec:3Daction}

Putting all the pieces together, the three-dimensional action  (\ref{eq:3Daction}) becomes: 
\begin{equation}
\begin{aligned}
\cL ~=~ & -\coeff{1}{4} \,e\,R ~-~  \coeff{1}{2}\,i e \, \overline{\chi}^{\dot{A} r}\gamma^\mu D_\mu\chi^{\dot{A} r} ~+~ \coeff{1}{2}  \,\epsilon^{\mu\nu\rho} \, \overline{\psi}^A_\mu D_\nu\psi^A_\rho   
 ~+~ \coeff{1}{8}\,e \, g^{\mu \nu} \, m^{IJ}   \, (\cD_\mu\, \chi_{I})  \, (\cD_\nu\, \chi_{J})   \\
 &~+~  \coeff{1}{16}\,e \, g^{\mu \nu} \,  \big( m^{IK} \, \cD_\mu\, m_{KJ}  \big)   \big( m^{JL} \, \cD_\nu\, m_{LI}  \big) ~+~  \coeff{1}{16}\,e \, g^{\mu \nu} \, \big( m^{IJ} \, m^{KL} \, C_{\mu}^{IK}  \, C_{\nu}^{JL}   \big)   \\
& ~+~  \coeff{1}{2}\,e  \, \varepsilon^{\mu \nu \rho} \, \Big[  g_0 \,\big(A_\mu{}^{IJ}\, \partial_\nu  \widetilde A_\rho{}^{IJ}  ~+~\coeff{4}{3}\,  g_0 \, A_\mu{}^{IJ} \,  A_\nu{}^{JK}\, A_\rho{}^{KI} \,\big) ~-~  \coeff{1}{4}\,  {B_\mu}{}^{IJ}  \, F_{\nu \rho}^{IJ} \Big]\\  
&~-~ 
\coeff{1}{2}\,e\, \cP_\mu^{Ir} \, \overline{\chi}^{\dot{A} r} \, \Gamma^I_{A\dot{A}}\gamma^\nu\gamma^\mu\psi^A_\nu
~+~ \coeff{1}{2}\,g\, e \, A_1^{AB} \,\overline{\psi}{}^A_\mu \gamma^{\mu\nu} \psi^B_\nu \\
&  
~+~  i\,g\,e \,A_2^{A\dot{A} r} \,\overline{\chi}^{\dot{A} r} \gamma^{\mu} \psi^A_\mu
~+~ \coeff{1}{2}\,g\,e \,A_3^{\dot{A} r\, \dot{B} s}\, \overline{\chi}^{\dot{A} r}\chi^{\dot{B} s}  ~-~e\, V \,.
\end{aligned}
\label{eq:3Daction2}
\end{equation}
One now completes the square in all the terms that involve ${B_\mu}{}^{IJ} $, to arrive at the action:
\begin{equation}
\begin{aligned}
\cL ~=~ & -\coeff{1}{4} \,e\,R ~-~  \coeff{1}{2}\,i e \, \overline{\chi}^{\dot{A} r}\gamma^\mu D_\mu\chi^{\dot{A} r} ~+~ \coeff{1}{2}  \,\epsilon^{\mu\nu\rho} \, \overline{\psi}^A_\mu D_\nu\psi^A_\rho   
 ~+~ \coeff{1}{8}\,e \, g^{\mu \nu} \, m^{IJ}   \, (\cD_\mu\, \chi_{I})  \, (\cD_\nu\, \chi_{J})   \\
 &~+~  \coeff{1}{16}\,e \, g^{\mu \nu} \,  \big( m^{IK} \, \cD_\mu\, m_{KJ}  \big)   \big( m^{JL} \, \cD_\nu\, m_{LI}  \big) ~-~   \coeff{1}{8}\, e \, g^{\mu \rho}  \, g^{\nu \sigma} \, m_{IK} \,m_{JL}\,  F_{\mu \nu }^{IJ}  \, F_{\rho \sigma }^{KL}   \\
& ~+~  \coeff{1}{2}\,e  \, \varepsilon^{\mu \nu \rho} \, \Big[  g_0 \,\big(A_\mu{}^{IJ}\, \partial_\nu  \widetilde A_\rho{}^{IJ}  ~+~\coeff{4}{3}\,  g_0 \, A_\mu{}^{IJ} \,  A_\nu{}^{JK}\, A_\rho{}^{KI} \,\big) ~+~  \coeff{1}{8}\,  {Y_\mu}{}^{IJ}  \, F_{\nu \rho}^{IJ} \Big]\\  
&~-~ 
\coeff{1}{2}\,e\, \cP_\mu^{Ir} \, \overline{\chi}^{\dot{A} r} \, \Gamma^I_{A\dot{A}}\gamma^\nu\gamma^\mu\psi^A_\nu
~+~ \coeff{1}{2}\,g\, e \, A_1^{AB} \,\overline{\psi}{}^A_\mu \gamma^{\mu\nu} \psi^B_\nu \\
&  
~+~  i\,g\,e \,A_2^{A\dot{A} r} \,\overline{\chi}^{\dot{A} r} \gamma^{\mu} \psi^A_\mu
~+~ \coeff{1}{2}\,g\,e \,A_3^{\dot{A} r\, \dot{B} s}\, \overline{\chi}^{\dot{A} r}\chi^{\dot{B} s}  ~-~e\, V ~+~ {\cal L}_{\rm B} \,,
\end{aligned}
\label{eq:3Daction3}
\end{equation}
where
\begin{equation}
\begin{aligned}
{\cal L}_{\rm B} ~\equiv~ &  \coeff{1}{16}\, e \, g^{\mu \nu} \, m^{IK} \,m^{JL}\,\Big({B_\mu}{}^{IJ}  +  \coeff{1}{2}\, Y_\mu{}_{ \, IJ}   - g_{\mu \sigma_1} \, \varepsilon^{\sigma_1  \rho_1 \rho_2} \, m_{I P_1} m_{J P_2}\,F_{\rho_1 \rho_2}^{P_1 P_2}\Big) \\
 &\qquad\qquad\qquad\qquad\qquad \times  \Big({B_\nu}{}^{KL} +  \coeff{1}{2} \,Y_\nu{}_{ \, KL} - g_{\nu \sigma_2} \, \varepsilon^{\sigma_2 \rho_3 \rho_4}\, m_{K P_3} \,m_{L P_4}\,F_{\rho_3 \rho_4}^{  P_3 P_4}   \Big) \,.
\end{aligned}
\label{Baction}
\end{equation}
The action  for the $B_\nu$ leads to the constraint:
\begin{equation}
 {C_\mu}{}^{IJ}~\equiv~  {B_\mu}{}^{IJ} ~+~ \coeff{1}{2}\,  Y_\mu{}_{ \, IJ} ~=~  g_{\mu \rho} \,  \varepsilon^{\rho \sigma \nu} \, m_{IK} m_{JL}\,F_{\sigma \nu}^{KL}   \,.
 \label{Beqn}
\end{equation}

These actions, and the constraint, are exactly consistent with the bosonic action given in \cite{Mayerson:2020tcl}.  One should note that here we are using a metric signature of $(+- -)$, whereas \cite{Mayerson:2020tcl} uses $(-++)$.  One can convert from one convention to the other by mapping $g_{\mu \nu} \to - g_{\mu \nu}$.  This reverses the sign of the Ricci scalar and, in (\ref{eq:3Daction2}) and  (\ref{Baction}), we have written the metric contractions explicitly so as to facilitate comparison.  One should remember that  $ e \, \varepsilon^{\mu \nu \rho}$ is actually independent of the metric  and so such terms remain unchanged under $g_{\mu \nu} \to - g_{\mu \nu}$.

\subsection{An executive summary of the BPS equations}
\label{sec:BPS}

For supersymmetric backgrounds we must require $\delta\psi^A_\mu   = \delta\chi^{\dot{A} r} = 0$.  These are the ``BPS equations'':
\begin{equation}
\delta\psi^A_\mu ~=~  D_\mu \epsilon^A ~-~ i  \,A_1^{AB}\gamma_\mu\epsilon^B ~=~ 0\,,    \qquad \delta\chi^{\dot{A} r} ~=~  \coeff{1}{2}\,i \,\Gamma^I_{A\dot{A}}\gamma^\mu\epsilon^A \, \cP_\mu^{Ir}  ~-~  A_2^{A\dot{A} r}\epsilon^A ~=~ 0\,.
\label{fermsusy}
\end{equation}
where we have taken $g =-1$ in accordance with (\ref{gchoice}).  Here we will simply summarize the pertinent details needed to set up and solve these equations.

First, the covariant derivative on $\epsilon^A$ is defined by:
\begin{equation}
D_\mu \, \epsilon^A ~=~    \partial_\mu \epsilon^A    ~+~ \coeff{1}{4}\,\omega_\mu{}^{ab}\,\gamma_{ab}\, \epsilon^A + \coeff{1}{4}\,\cQ_\mu^{IJ}\Gamma^{IJ}_{AB}\, \epsilon^B \,,   
\label{Depsilon}
\end{equation}
where
\begin{equation}
\cQ_\mu^{IJ}   ~=~  \coeff{1}{2}\, \Big[ \, \big(P^{-1} \cD_\mu \, P\big)_{I}{ }^{\, J}  ~-~  \big(P^{-1} \cD_\mu \, P\big)_{J}{}^{\, I} ~+~\scrC_{\mu}^{IJ} \   \,  \Big] \,. 
\label{Qdefn1}
\end{equation}
and $P= P^T$ is a symmetric $GL(4,\IR)$ matrix.  The gauge covariant derivative is:
\begin{equation}
\cD_\mu \, \cX_{I}  ~=~ \partial_\mu \, \cX_{I}   ~-~  2\, g_0\,\widetilde A_\mu{}^{IJ} \, \cX_{J} \,. 
\label{covderiv5}
\end{equation}
and it acts only on the left-hand side of $P$.  The Chern-Simons vector fields are determined by:
\begin{equation}
\scrC_{\mu}^{IJ}   ~\equiv~  P^{-1}{}_{I}{}^K \,  P^{-1}{}_{J}{}^L \, C_\mu^{KL}\,, \qquad   {C_\mu}{}^{IJ}~=~    g_{\mu \rho} \,  \varepsilon^{\rho \sigma \nu} \, m_{IK} m_{JL}\,F_{\sigma \nu}^{KL}  \,.
 \label{scrCdefn}
\end{equation}
The scalar kinetic terms, $\cP_\mu^{Ir}$,  are given by:
\begin{equation}
\begin{aligned}
\cP_\mu^{Ir}   ~=~ & \coeff{1}{2}\, \Big[ \, \big(P^{-1} \cD_\mu \, P\big)_{I}{ }^{\, r}  ~+~  \big(P^{-1} \cD_\mu \, P\big)_{r}{}^{\, I} ~-~\scrC_{\mu}^{Ir} \   \,  \Big] \,, \qquad 1 \le r  \le 4 \,, \\
\cP_\mu^{I5}   ~=~&  - \coeff{1}{\sqrt{2}}\, \big(P^{-1} \big)_{I}{ }^{\, A}  \,\cD_\mu  \chi_A \,,  \qquad 1 \le r  \le 4  \,.
 \end{aligned}
\label{ScalarKin2}
\end{equation}
We take the scalar matrix to be diagonal:
\begin{equation}
P  ~=~  {\rm diag} \big(\,  e^{\mu_1} \,, \,  e^{\mu_2} \,, \,  e^{\mu_3} \,, \,  e^{\mu_4} \, \big) \,.
\label{Pdiag2}
\end{equation}
The superpotential is defined by:
\begin{equation}
\begin{aligned}
W  ~\equiv~ & \coeff{1}{4} \, g_0 \,  (\det(P))^{-1}  \,   \Big [\, 2 \,\Big(1- \coeff{1}{4} \,  (\chi_A \chi_A)\Big) ~-~ {\rm Tr}\big(P\, P^T\big)  \, \Big] \\
~=~ & \coeff{1}{4} \, g_0 \,e^{-\mu_1 -\mu_2-\mu_3-\mu_4} \, \Big [\, 2 \,\Big(1- \coeff{1}{4} \,  (\chi_A \chi_A)\Big) ~-~ \Big( e^{2\, \mu_1}+e^{2\, \mu_2}+e^{2\, \mu_3}+e^{2\, \mu_4}  \Big) \, \Big]   \,,
 \end{aligned}
\label{superpot2}
\end{equation}
and the $A$-tensors are given by:
\begin{equation}
\begin{aligned}
A_1^{AB}   ~=~ W\, (\Gamma^{1234}\big)_{AB} \,, \qquad A_2^{A\dot A \, r}   ~=~ & \frac{\partial W}{\partial \mu_r} \, (\Gamma^{1234}\,\Gamma^{r}\big)_{A \dot A} \,, \quad 1 \le r \le 4  \,, \\
\qquad A_2^{A\dot A \, 5}   ~=~ &- \sqrt{2}\, \sum_{j=1}^4 \,e^{\mu_j} \, \frac{\partial W}{\partial \chi_j} \, (\Gamma^{1234}\,\Gamma^{j}\big)_{A \dot A}  \,,
\end{aligned}
\label{A1A2tens2}
\end{equation}
where there is no sum on $r$ in the expression for $A_2^{A\dot A \, r}$.

\section{Superstrata in three dimensions}  
\label{sec:superstrata_in_3_dimensions}

In  \cite{Mayerson:2020tcl} it was shown how to reduce the $(1,m,n)$ family of superstrata to an entirely three-dimensional description.  In this section we will   summarize these results and use them to compute all the individual terms that go into the three-dimensional BPS equations.   In the next Section, we will use all this data  to solve the BPS equations.

\subsection{The metric}
\label{sec:thessmet}

Following on from Section \ref{sec:metric}, we use the coordinates $(u,v,r)$ and work with asymptotically AdS geometries and use $\xi$ defined in (\ref{xidef}).  For the superstrata, the three dimensional metric has the form \cite{Mayerson:2020tcl}:
\begin{equation}
    ds_3^2 ~=~ \frac{a^4 R_y^2 g_0^6}2 \qty(\dd{u} + \dd{v} + \frac{\sqrt{2}}{a^2 R_y g_0^4} \, \mathcal{A})^2 ~-~   \frac{\Lambda^2}{g_0^2} \, ds_2^2  \,,
\end{equation}
where
\begin{equation}
    ds_2^2  ~=~  \frac{\abs{\dd{\xi}}^2}{\qty(1-\abs{\xi}^2)^2} \qand \mathcal{A} ~=~ \frac i2 \frac{\xi \dd\bar\xi - \bar\xi\dd\xi}{1- \abs{\xi}^2} ~=~   \frac{\sqrt{2}\,r^2}{a^2 R_y} \dd{v} \,.
 \label{2dbasemet}
\end{equation}
Note that, compared to (\ref{genmet}), there is only one arbitrary function, $\Lambda$, in this metric.  In particular, the ``time fibration'' part of the metric is a scaled version of that of the AdS metric.

For future reference, we note that the metric, $ds_2^2$, has a K\"ahler potential
\begin{equation}
\mathscr{K}  ~\equiv~  - \log \big( 1 - |\xi|^2 \big) 
\label{Kahlerpot}
\end{equation}
and that $\cA$ is the potential for the K\"ahler form. Thus the three-dimensional metric has the form of a canonical time-like K\"ahler fibration.

As we will see, supersymmetry requires that this warp factor be fixed in terms of the scalars:
\begin{equation}
   \Lambda^2  ~=~ 1 ~-~ \coeff{1}{4} \, \big( \chi_1^2 + \chi_2^2 +  \chi_3^2 + \chi_4^2  \big)   \,.
    \label{Lambdaform}
\end{equation}

We will use the frames: 
\begin{equation}
e^0 ~=~ \frac{a^2 R_y g_0^3}{\sqrt{2}} \qty(\dd{u} + \dd{v}) + \frac{1}{g_0} \mathcal{A} \,,  \qquad 
e^1 ~=~ \frac {\Lambda}{g_0}\ \frac{1}{\sqrt{r^2 + a^2}} \dd{r}\,, \qquad
e^2 ~=~ \frac{\Lambda}{g_0}\ \frac{\sqrt{2}\ r\, \sqrt{r^2 + a^2}}{a^2 R_y} \dd{v}\,,
\label{frames}
\end{equation}
such that
\begin{equation}
    ds_3^2 ~=~ \qty(e^0)^2 - \qty(e^1)^2 - \qty(e^2)^2 \ .
\end{equation}

The spin connection is then given by
\begin{equation}
    \begin{aligned}
        \omega\indices{^0_1} &~=~   \omega\indices{^1_0}~=~   \frac{g_0}{\Lambda^2} \,e^2 \,, \qquad   \omega\indices{^0_2}~=~   \omega\indices{^2_0} ~=~ - \frac{g_0}{\Lambda^2} \,e^1 \,, \\
        \omega\indices{^1_2} &~=~  - \omega\indices{^2_1}~=~ \frac{g_0}{\Lambda^2} \,e^0 \,+\,\frac{\lambda_2}{2 \Lambda^2}\, e^1 \,-\, \frac{\lambda_1}{2 \Lambda^2} \,e^2 \,-\, \frac{g_0}{\Lambda}  \, \frac{2 \,r^2 + a^2}{r\, \sqrt{r^2 + a^2}}  \,e^2\,,
    \end{aligned}
    \label{spin_connection}
\end{equation}
where $\lambda_1$ and $\lambda_2$ are defined by:
\begin{equation}
    \dd(\Lambda^2) ~\equiv~  \lambda_1 \, e^1 + \lambda_2  \, e^2 \,.
    \label{lambda_der}
\end{equation}
%

\subsection{The scalars}
\label{sec:ssscalars}

The fundamental scalars that determine the superstratum fluxes are parameterized by two holomorphic functions, $F_0$ and $F_1$, of $\xi$: 
\begin{equation}
\begin{aligned}
        \chi_1 + i \chi_2 ~=~ & -\frac{2\sqrt{2} \,a \, R_y \,g_0}{\sqrt{r^2 + a^2}}\ i\, F_0 (\xi)   ~=~  -2\sqrt{2} \,a R_y g_0 \, i\, \Big(
e^{-\frac{1}{2}\,  \mathscr{K} }\ F_0 (\xi)\Big) \,, \\ 
        \chi_3 - i \chi_4 ~=~ & \frac{2\sqrt{2} \,a \, R_y \,g_0}{\sqrt{r^2 + a^2}}\ i\, F_1 (\xi) ~=~  2\sqrt{2} \,a R_y g_0\, i\, \Big( e^{-\frac{1}{2}\,  \mathscr{K} } F_1 (\xi) \Big) \,.
\end{aligned}
 \label{chi_superstrata}
\end{equation}
where we have written these scalars in a more canonical form using the K\"ahler potential (\ref{Kahlerpot}).

One should note that our expression for $ \chi_3 - i \chi_4 $ differs by a phase from that of  \cite{Mayerson:2020tcl}.  We have performed a $U(1)$ gauge transformation so as to make  $\chi_3 - i \chi_4 $ have the same form as   $\chi_1 + i \chi_2 $.  As we will see, this gauge transformation also makes slight modifications elsewhere.   One should also note the difference of sign in the two left-hand sides  of (\ref{chi_superstrata}): this will play a crucial role in the supersymmetry.

For future reference we note that the $(1,0,n)$ family of superstrata is defined by taking $ \chi_3 - i \chi_4 =0$ and the $(1,1,n)$ family is defined by taking $ \chi_1 +  i \chi_2 =0$.

To describe the scalar sector, it is convenient to introduce the shorthand:
\begin{equation}
    \rho_1^2 ~=~ \chi_1^2 + \chi_2^2 \ , \quad \rho_2^2 ~=~ \chi_3 ^2 + \chi_4^2 \ , \quad \rho_0^2 ~=~ \rho_1^2 + \rho_2^2 \qand \Lambda^2 ~=~ 1 - \coeff{1}{4}\, \rho_0^2 \,.
    \label{rho_defs}
\end{equation}

The  scalar matrix $m$, with components $m_{IJ}$,  which descends from the shape modes on $S^3$, is given by:
\begin{equation}
    m = \oneone ~-~  \frac 14 \mqty(
    \rho_1^2 & 0        & \chi_1 \chi_3 - \chi_2 \chi_4   & \chi_1 \chi_4 + \chi_2 \chi_3 \\
    0        & \rho_1^2 & \chi_1 \chi_4 + \chi_2 \chi_3     & -(\chi_1 \chi_3 - \chi_2 \chi_4) \\
    \chi_1 \chi_3 - \chi_2 \chi_4   & \chi_1 \chi_4 + \chi_2 \chi_3     & \rho_2^2 & 0 \\
    \chi_1 \chi_4 + \chi_2 \chi_3     & -(\chi_1 \chi_3 - \chi_2 \chi_4)  & 0        & \rho_2^2) \,.
\label{genmmat}
\end{equation}
Note that this matrix is diagonal for the $(1,0,n)$ and $(1,1,n)$ sub-families separately.    We also note that, while this matrix has exactly the same functional form as that of  \cite{Mayerson:2020tcl}, it is, in fact, different.  This is because we have made a gauge transformation to remove a phase from  $\chi_3 - i \chi_4$.  This gauge transformation also acts on the matrix $m$ and preserves its functional form despite the non-trivial change in  $\chi_3 - i \chi_4$.

\subsection{The gauge fields}
\label{sec:ssgauge}

The  gauge fields live in an $SU(2) \times U(1)$ subgroup of $SO(4)$ and so we introduce the matrices:
\begin{equation}
\eta_1   ~\equiv~  
\left( \begin{matrix} 
0  & 0& 0 & 1 \\
0  & 0& 1 & 0  \\
0  & -1& 0 & 0  \\
-1  & 0& 0 & 0  \\
\end{matrix} \right) \,, \qquad 
\eta_2  ~\equiv~  
\left( \begin{matrix} 
0  & 0& -1 & 0 \\
0  & 0& 0 & 1 \\
1  & 0& 0 & 0  \\
0  & -1& 0 & 0  \\
\end{matrix} \right)  \,,  \qquad  
\eta_3  ~\equiv~  
\left( \begin{matrix} 
0  & 1& 0 & 0 \\
-1 & 0& 0 & 0  \\
0  & 0& 0 & 1 \\
0  & 0& -1  & 0  \\
\end{matrix} \right) \,, 
\label{etamat}
\end{equation}
\begin{equation} 
\bar\eta_1  ~\equiv~  
\left( \begin{matrix} 
0  & 0& 0 & 1 \\
0  & 0& -1 & 0  \\
0  & 1& 0 & 0  \\
-1  & 0& 0 & 0  \\
\end{matrix} \right) \,, \qquad 
\bar\eta_2  ~\equiv~  
\left( \begin{matrix} 
0  & 0& -1 & 0 \\
0  & 0& 0 & -1 \\
1  & 0& 0 & 0  \\
0  &  1& 0 & 0  \\
\end{matrix} \right)  \,,   \qquad  
\bar\eta_3  ~\equiv~  
\left( \begin{matrix} 
0  & 1& 0 & 0 \\
-1 & 0& 0 & 0  \\
0  & 0& 0 & -1 \\
0  & 0& 1  & 0  \\
\end{matrix} \right)    \,.
\label{etabarmat}
\end{equation}
 The triplet $(\eta_1, \eta_2, \eta_3)$ generates one of $\mathfrak{su}(2)$ algebra, while the triplet $(\bar\eta_1, \bar\eta_2, \bar\eta_3)$ generates the other commuting $\mathfrak{su}(2)$ algebra.   

The gauge fields  are then given by:
\begin{equation}
    \tilde A^{IJ} ~=~ \frac1{\sqrt{2}a^2 R_y g_0} \qty(C_1 \eta_1^{IJ} + C_2 \eta_2^{IJ} + C_3 \eta_3^{IJ} + \bar C_3 \bar\eta_3^{IJ})
 \label{gengauge}
\end{equation}
where
\begin{align}
    C_1 ~=~ &   \coeff{1}{4} \, \big(\chi_1 \chi_3 - \chi_2 \chi_4  \big)\, \nu \, \qquad 
    C_2 ~=~ \coeff{1}{4} \, \big( \chi_1 \chi_4 + \chi_2 \chi_3 \big)\, \nu  \, \qquad  
    C_3 ~=~ - \coeff{1}{8}\,  \qty(\rho_1^2 - \rho_2^2)\,  \nu \,,\\
    \bar C_3 ~=~ & - r^2 \dd{v} + \big(1- \coeff{1}{8} \,\big(\rho_1^2 + \rho_2^2\big) \big) \, \nu \,,
\end{align}
and
\begin{equation}
    \nu   ~\equiv~  \frac 1{\Lambda^2} \qty( \frac{a^4 R_y^2 \, g_0^4}{2} (\dd{u} + \dd{v}) + r^2 \dd{v}) ~=~  \frac{a^2 R_y\, g_0}{\sqrt{2} \, \Lambda^2} \, e^0 \,.
\end{equation}
Once again, these gauge fields are slightly different from those given in \cite{Mayerson:2020tcl} because we have made a gauge transformation to remove a phase from  $\chi_3 - i \chi_4$.

\subsection{The gauge fields for the $(1,0,n)$ superstratum}
\label{sec:gauge10n}

It is extremely instructive to consider the pure  $(1,0,n)$ superstratum in which one has $\chi_3 =\chi_4 =0$.  The scalar matrix reduces to:  
\begin{equation}
    m ~=~  \mqty(\dmat{\Lambda^2, \Lambda^2, 1, 1}) \qquad \Rightarrow  \qquad P  ~=~  \mqty(\dmat{\Lambda , \Lambda, 1, 1})  \,,
    \label{u1xu1_m}
\end{equation}
and the gauge fields (\ref{gengauge}) become
\begin{equation}
    \begin{split}
        \tilde{A}^{12} &~=~ \frac{a^2 R_y \, g_0^3}{2 \sqrt{2}} \, (\dd{u} + \dd{v}) ~=~ \frac{e^0}2 ~-~ \frac{r}{\sqrt{r^2 + a^2}} \frac{e^2}{2\Lambda} \,, \\
        \tilde{A}^{34} &~=~ -\frac{a^2 R_y \, g_0^3}{2 \sqrt{2} \Lambda^2}  \, (\dd{u} + \dd{v}) ~+~ \frac{\Lambda^2 -1}{\Lambda^2} \frac{r^2}{\sqrt{2}a^2 R_y g_0} \,  \dd{v} ~=~ - \frac{e^0}{2 \Lambda^2} ~+~ \frac{r}{\sqrt{r^2 + a^2}} \, \frac{e^2}{2\Lambda} \,.
    \end{split}
\label{u1xu1_gauge}
\end{equation}
Since this connection is abelian, the field strength is simply $F = \dd A$ and its components are given by:
\begin{equation}
F^{12} ~=~ - \frac{1}{\Lambda^4} \,\Big[ \,g_0 \,  ( 1- \Lambda^2)\,  e^1 \wedge e^2 ~+~ \coeff{1}{2}\, e^0 \wedge \big(\lambda_1 e^1 + \lambda_2 e^2\big)  \, \Big] \,,  \qquad 
F^{34}~=~ 0\ .
    \label{u1xu1_strength}
\end{equation}
We then find that  the components of the Chern-Simons terms,  (\ref{cs_definition}), are given by:
\begin{equation}
\mathscr{C}^{12} ~=~ - \frac{1}{\Lambda^2} \,\Big[ 2 g_0 \,(1-\Lambda^2)\, e^0 ~+~  \lambda_2 \,  e^1 ~-~ \lambda_1 \, e^2 \Big] \,, \qquad 
\mathscr{C}^{34} ~=~ 0 \ .
\label{u1xu1_cs}
\end{equation}
Finally, the scalar kinetic terms (\ref{ScalarKin2}) are:
\begin{equation}
\begin{aligned}
\cP^{Ir}   ~=~ & \frac{1}{2\Lambda^2}\, \big( \, \lambda_1 \, e^1 + \lambda_2  \, e^2\,\big)\,  \delta_I^r ~-~ \coeff{1}{2}\,  \scrC^{Ir} \ , \qquad 1 \le r  \le 2 \,, \\
\cP_\mu^{I3} ~=~ & \cP_\mu^{I4} ~=~ 0 \,, \qquad 
\cP_\mu^{I5}   ~=~  - \frac{1}{\sqrt{2}\, \Lambda}\,\qty(\cD_\mu \chi)_I  \,.
 \end{aligned}
\label{u1xu1_scalar_kin}
\end{equation}
and the connection $\cQ_\mu^{IJ}$, defined in (\ref{Qdefn1}), becomes
\begin{equation}
\cQ_\mu^{IJ} ~=~ - 2\, g_0 \, A_\mu^{IJ}  ~+~ \coeff{1}{2}\,  \scrC_\mu^{IJ} \,.
\label{Qsimp1}
\end{equation}
%

\subsection{A gauge transformation of the full $(1,m,n)$ superstratum}
\label{sec:gauge1mn}

We have written the supersymmetry transformations in terms of the ``diagonal gauge'' for $P$, (\ref{Pdiag2}), however (\ref{genmmat}) is not in that gauge.  One can either recast the supersymmetry transformations in a general gauge, or one can diagonalize $m$.  We choose the latter option.

Define
\begin{equation}
\mathscr{U}
~\equiv~
\left( \begin{matrix} 
\frac{\rho_1}{\rho_0} & 0 &\frac{1}{\rho_0\, \rho_1} \,(\chi_1 \chi_3 - \chi_2 \chi_4)  &\frac{ 1}{\rho_0\, \rho_1}\, (\chi_1 \chi_4 + \chi_2 \chi_3)   \\
0 & \frac{\rho_1}{\rho_0} & \frac{ 1}{\rho_0\, \rho_1}\, (\chi_1 \chi_4 + \chi_2 \chi_3) & -\frac{1}{\rho_0\, \rho_1} \,(\chi_1 \chi_3 - \chi_2 \chi_4)  \\
0 & \frac{\rho_2}{\rho_0}  & -\frac{ 1}{\rho_0\, \rho_2}\, (\chi_1 \chi_4 + \chi_2 \chi_3) & \frac{1}{\rho_0\, \rho_2} \,(\chi_1 \chi_3 - \chi_2 \chi_4) \\
\frac{\rho_2}{\rho_0} & 0 & -\frac{1}{\rho_0\, \rho_2} \,(\chi_1 \chi_3 - \chi_2 \chi_4) & -\frac{ 1}{\rho_0\, \rho_2}\, (\chi_1 \chi_4 + \chi_2 \chi_3)   \\
\end{matrix} \right) \,,
\label{diagonalization}
\end{equation}
This is an $SO(4)$ matrix.  Indeed, it commutes with $ \bar\eta_3^{AB}$, and so lies in the same $SU(2) \times U(1)$ as the gauge connection (\ref{gengauge}).  By construction, one has:
\begin{equation}
 \hat m ~\equiv~ \mathscr{U} \, m  \,\mathscr{U}^{-1} ~=~    \mqty(\dmat{\Lambda^2, \Lambda^2, 1, 1}) \qquad \Rightarrow  \qquad \hat P ~=~ \mqty(\dmat{\Lambda , \Lambda, 1, 1})\,,
 \label{DiagmP}
\end{equation}
where $\Lambda$ is defined by (\ref{Lambdaform}).   One also finds that
\begin{equation}
 (\hat\chi_1, \hat\chi_2, \hat\chi_3, \hat\chi_4) ~\equiv~     \mathscr{U} \,  (\chi_1, \chi_2, \chi_3, \chi_4) ~=~  \frac{\rho_0}{\rho_1}\, \qty( \chi_1,  \chi_2, 0, 0)  \,.
 \label{trfchi}
\end{equation}
Observe that, up to an overall factor, $\hat\chi_i$ is the same as for of the $(1,0,n)$ superstratum.  For future analysis, it is useful to separate out this factor and  define:
\begin{equation}
\tilde \chi_I ~\equiv~   \frac{\rho_1}{\rho_0}\,    \hat\chi_I \,, \qquad \Rightarrow \qquad \tilde \chi ~=~  \qty( \chi_1,  \chi_2, 0, 0)  \,.
 \label{tildechi}
\end{equation}
One should also note that 
\begin{equation}
 \hat \Lambda^2 ~\equiv~    1 ~-~ \coeff{1}{4} \, \big( \hat \chi_1^2 + \hat \chi_2^2 +  \hat \chi_3^2 + \hat \chi_4^2  \big)  ~=~\Lambda^2 \,,
 \label{hatLamda}
\end{equation}
is gauge invariant.

The gauge transformation also significantly simplifies the gauge field:
\begin{equation}
\mathscr{U} \, \tilde A  \,\mathscr{U}^{-1} ~=~ \frac{1}{2\,\sqrt{2} \, R_y a^2 g_0}  \, \Big[\, \big(\Lambda^2 \, (\eta_3 +  \bar\eta_3)   +( \bar\eta_3 -\eta_3 ))\, \nu 
 - 2\, r^2 \, \bar\eta_3 \dd{v} \, \Big]  ~\equiv~  \tilde A_\text{Abelian} 
 \label{AAbeldefin}\,. 
\end{equation}
In terms of components, this implies that the only non-zero components are:
\begin{equation}
\begin{aligned}
\tilde A_\text{Abelian}^{12} ~=~ \big(\mathscr{U} \, \tilde A  \,\mathscr{U}^{-1}\big)^{12} ~=~ & \frac{e^0}{2} ~-~ \frac{r}{\sqrt{r^2 + a^2}} \frac{e^2}{2\Lambda} \,,  \\  
\tilde A_\text{Abelian}^{34} ~=~\big(\mathscr{U} \, \tilde A  \,\mathscr{U}^{-1}\big)^{34} ~=~ &- \frac{e^0}{2 \Lambda^2} ~+~ \frac{r}{\sqrt{r^2 + a^2}} \, \frac{e^2}{2\Lambda}   \,,
\end{aligned}
\label{AAbelparts}
\end{equation}
which exactly matches the gauge connection (\ref{u1xu1_gauge}) for the $(1,0,n)$ superstratum.   We have thus almost mapped the complete superstratum back onto the $(1,0,n)$ superstratum using the local $SU(2) \times U(1)$ gauge transformation defined by $\mathscr{U}$.   There are, however, two important differences.  First, the functional dependence of  (\ref{trfchi}) is a little more complicated than that of (\ref{chi_superstrata}), and second, the transformed gauge potential is, of course:
\begin{equation}
\hat {\tilde A} ~=~ \tilde A_\text{Abelian} ~+~ \frac{1}{2 \, g_0} \,  \big(\dd \mathscr{U}   \big) \,\mathscr{U}^{-1} \,.
\label{Ahatdefn}
\end{equation}

To write the last term in (\ref{Ahatdefn}), we define:
\begin{equation}
\begin{aligned}
 K_1 ~\equiv~ & \frac{1}{\rho_1^2}\,(\chi_2 \dd{\chi_1} - \chi_1 \dd{\chi_2}) ~=~  \dd \arctan \bigg(\frac{\chi_1}{\chi_2} \bigg)   \,, \\
    K_2 ~\equiv~& \frac{1}{\rho_2^2}\,(\chi_4 \dd{\chi_3} - \chi_3 \dd{\chi_4}) ~=~  \dd \arctan \bigg(\frac{\chi_3}{\chi_4} \bigg)\,, \\
  L_1 ~\equiv~ & \dd \log(\rho_1)   \,, \qquad L_2 ~\equiv~ \dd \log(\rho_2)  \,,
\end{aligned}
\label{KLdefn}
\end{equation}
and then one has
\begin{equation}
\big(\dd \mathscr{U}   \big) \,\mathscr{U}^{-1} ~=~ \frac{1}{2\, \rho_0^2} \, (K_1 + K_2) \,\Big[ (\rho_2^2 -\rho_1^2 ) \, \eta_3 +  \rho_0^2 \, \bar\eta_3 ~+~ 2\, \rho_1 \rho_2\,\eta_2 \,\Big] ~+~  \frac{\rho_1 \rho_2}{\rho_0^2} \big(L_1 - L_2\big)\, \eta_1 \,.
\label{puregauge}
\end{equation}

The field strength is then given by: 
\begin{equation}
    \hat {\tilde F}  ~=~ \tilde F_\text{Abelian}  ~+~  \frac{\rho_1\rho_2}{8\,\Lambda^2} \,  e^0 \wedge \big( \, (K_1 + K_2)\, \eta_1 ~-~ (L_1-L_2)\, \eta_2\, \big)
\end{equation}
where $\tilde F_\text{Abelian} = \dd \tilde A_\text{Abelian} $ and   $\tilde A_\text{Abelian} $ is defined in  (\ref{AAbeldefin}), or  (\ref{u1xu1_gauge}).  After taking the $SO(4)$ dual,  $F_\text{Abelian}$ can be read off from  (\ref{u1xu1_strength}).

To compute the Chern-Simons terms we need the frame components of the $K$'s and $L$'s, and so we write 
\begin{equation}
 (K_1 + K_2)  ~=~   \cK_1 \, e^1 ~+~ \cK_2 \, e^2 \,, \qquad  (L_1 - L_2)  ~=~  \cL_1 \, e^1 ~+~\cL_2 \, e^2 \,.
 \label{cKcLdefn}
\end{equation}
We then obtain:
\begin{equation}
\begin{aligned}
 \hat{\mathscr{C}}  ~=~& - \frac{1}{2\, \Lambda^2} \,\Big[ \, 2 g_0 \,(1-\Lambda^2)\, e^0 ~+~  \lambda_2 \,  e^1 ~-~ \lambda_1 \, e^2 \, \Big] \, (\eta_3 + \bar\eta_3)  \\
& ~+~  \frac{\rho_1\rho_2}{4 \, \Lambda} \,\Big[ \, \big(  \cK_2 \,  e^1 ~-~ \cK_1 \, e^2 \big)\, \eta_1 ~-~ \big(  \cL_2 \,  e^1 ~-~ \cL_1 \, e^2 \big)\, \eta_2     \, \Big]   \,.
\end{aligned}
\label{fullCSterm}
\end{equation}
It is convenient to define the ``Abelian'' piece of this connection:
\begin{equation}
\mathscr{C}_\text{Abelian} ~\equiv~  - \frac{1}{2\, \Lambda^2} \,\Big[ \, 2 g_0 \,(1-\Lambda^2)\, e^0 ~+~  \lambda_2 \,  e^1 ~-~ \lambda_1 \, e^2 \, \Big] \, (\eta_3 + \bar\eta_3) 
  \,,
\label{CAbelian}
\end{equation}
and we note that this is exactly the Chern-Simons connection (\ref{u1xu1_cs}) for the $(1,0,n)$ superstratum.

Finally, the scalar kinetic terms are a little more complicated than those of (\ref{u1xu1_scalar_kin}).  We must  use  $\hat P$, defined in  (\ref{DiagmP}), which is identical to $P$ in   (\ref{u1xu1_m}), in (\ref{ScalarDers}).   The difference now is that the gauge field, $\tilde A$, is no longer $U(1) \times U(1)$ invariant and so $\cP^{Ir}$ and $\cQ^{IJ}$ have new gauge terms.    We find:
\begin{equation}
\begin{aligned}
\cP^{Ir}   ~=~ &  \dd \log(\Lambda) \,  \delta_I^r ~-~ \frac{\rho_1\rho_2}{8  \, \Lambda} \,\Big[ \,(L_1 - L_2)\, \eta_1 ~+~(K_1 + K_2)\, \eta_2     \, \Big]   ~-~ \coeff{1}{2}\,  \scrC^{Ir} \ , \qquad 1 \le r  \le 2 \,, \\
\cP^{Ir}   ~=~ & + \frac{\rho_1\rho_2}{8  \, \Lambda} \,\Big[ \,(L_1 - L_2)\, \eta_1 ~+~(K_1 + K_2)\, \eta_2     \, \Big]   ~-~ \coeff{1}{2}\,  \scrC^{Ir} \ , \qquad 3 \le r  \le 4 \,, \\
\cP_\mu^{I5}   ~=~ & - \frac{1}{\sqrt{2}\, \Lambda}\,\qty(\cD_\mu \chi)_I  \,.
 \end{aligned}
\label{genscalar_kin}
\end{equation}
and the connection $\cQ_\mu^{IJ}$, defined in (\ref{Qdefn1}), becomes
\begin{equation}
\cQ_\mu^{IJ} ~=~ 
\begin{cases}
- 2\, g_0 \, \tilde A_\mu^{IJ}  ~+~ \coeff{1}{2}\,  \scrC_\mu^{IJ}  & I,J \in \{1,2\} \ {\rm or} \   I,J \in \{3,4\}  \\  
- 2\, g_0 \, \frac{1 + \Lambda^2}{\Lambda}   \, \tilde A_\mu^{IJ}  ~+~ \coeff{1}{2}\,  \scrC_\mu^{IJ}  & I \in \{1,2\}\,,  J \in \{3,4\} \ {\rm or} \ I \in \{3,4\}  \,,    J \in \{1,2\}
\end{cases}
\,.
\label{QIJfinal}
\end{equation}
Observe how the signs in front of the $\eta$-matrix terms flip between the first and second line  of (\ref{genscalar_kin}).  This happens because because the covariant derivatives of $P$ are symmetrized in $\cP^{Ir}$.

\subsection{Holomorphy}
\label{sec:holomorphy}

There are many significant aspects to holomorphy in the structure of the superstrata, but for now we focus on how this influences the solution of the BPS equations.  In particular, we first observe that all the scalar fields, $\chi_i$, in (\ref{chi_superstrata}) involve a common, non-holomorphic pre-factor of $(r^2 + a^2)^{-\frac{1}{2}}$.  (As we saw in (\ref{chi_superstrata}), this factor has a natural interpretation in terms of the K\"ahler potential.)  The derivatives of this factor cancel out in $(K_1 + K_2)$ and $(L_1 - L_2)$, leaving only the derivatives of the holomorphic functions, $F_0$ and $F_1$.   

Since we are working in real coordinates, it is simplest to express the holomorphy properties  in terms of the Cauchy-Riemann equations, which take a very simple form  when expressed in terms of the frames, $e^1$ and $e^2$.  In particular,  because the $\chi_i$'s are holomorphic up to a  common pre-factor, we find that the Cauchy-Riemann equations imply:
\begin{equation}
 \cK_1  ~=~     \cL_2 \,, \qquad     \cK_2  ~=~    - \cL_1 \,.
 \label{relationKL}
\end{equation}

One consequence of this is that the Chern-Simons term (\ref{fullCSterm}) can be re-written as:
\begin{equation}
 \hat{\mathscr{C}} ~=~ \mathscr{C}_\text{Abelian}  
 ~-~  \frac{\rho_1\rho_2}{4 \, \Lambda} \,\Big[ \,(L_1 - L_2)\, \eta_1 ~+~(K_1 + K_2)\, \eta_2     \, \Big]   \,,
\label{fullCSterm2}
\end{equation}
where $\mathscr{C}_\text{Abelian}$ is defined in (\ref{CAbelian}).

As a result, we find that various pieces of the gauge connection and the Chern-Simons connection either cancel, or reinforce, in the $4 \times 4$ block of the scalar kinetic term:
\begin{equation}
\cP^{IJ}   ~=~ 
\begin{cases}
 \dd \log(\Lambda) \,  \delta^{IJ}   ~-~ \coeff{1}{2}\,  \scrC^{IJ}\qquad\qquad     & I,J \in \{1,2\}  \\[1ex]
0   & I \in \{1,2\}\,,  J \in \{3,4\} \\
- \scrC^{IJ}   &   I \in \{3,4\}   \,,  J \in \{1,2\}  \\
0    & I,J \in \{3,4\} 
\end{cases}
\,,
\label{PIJfinal}
\end{equation}
and $\cP_\mu^{I5}$ is unmodified:
\begin{equation}
\cP_\mu^{I5}   ~=~  - \frac{1}{\sqrt{2}\, \Lambda}\,\qty(\cD_\mu \chi)_I  \,.
\label{Pi5same}
\end{equation}

Holomorphy thus plays a critical role in the cancellation that produces the second row of (\ref{PIJfinal}), and, as we will see, this is essential to the supersymmetry.

Holomorphy also leads to another  important identity.  Observe that if $\chi_1$ and $\chi_2$ have the form (\ref{chi_superstrata}) then $(\dd - i \cA) (\chi_1 + i \chi_2)$ is a holomorphic differential, and thus proportional to $e^1 + i e^2$. This follows because the anti-holomorphic differentials cancel between $\cA$ and $\dd \mathscr{K}$.  
If one writes the differentials in terms of the real frame components: 
\begin{equation}
\begin{aligned}
\dd \chi_1  +  \cA \, \chi_2    ~\equiv~ (\dd \chi_1+  \cA \, \chi_2)_1\, e^1  ~+~ (\dd \chi_1  +  \cA \, \chi_2)_2\, e^2      \,, \\
\dd \chi_2  -  \cA \, \chi_1    ~\equiv~ (\dd \chi_2  -  \cA \, \chi_1 )_1\, e^1  ~+~ (\dd \chi_2  -  \cA \, \chi_1 )_2\, e^2      \,,
\end{aligned}
\label{holdiffchi}
\end{equation}
then holomorphy implies the Cauchy-Riemann conditions:
\begin{equation}
 (\dd \chi_1+  \cA \, \chi_2)_1  ~=~  (\dd \chi_2  -  \cA \, \chi_1 )_2  \,, \qquad     (\dd \chi_1  +  \cA \, \chi_2)_2  ~=~ -  (\dd \chi_2  -  \cA \, \chi_1 )_1  \,.
\label{chiCR}
\end{equation}

Note that  (\ref{frames})  and (\ref{AAbelparts}) imply
\begin{equation}
\tilde A_\text{Abelian}^{12} ~=~ \frac{e^0}{2} ~-~ \frac{r}{\sqrt{r^2 + a^2}} \frac{e^2}{2\Lambda} ~=~ \frac{1}{2}\,e^0 ~-~ \frac{1}{2\,g_0 } \cA\,,  
\label{AAbelform1}
\end{equation}
which means that the covariant derivatives of $\chi_1$ and $\chi_2$ contain precisely the terms that are related by (\ref{chiCR}).

There is a  parallel story for $\chi_3 - i \chi_4$  if this also has the form given in (\ref{chi_superstrata}).

Finally, we note that while the scalar matrix, $m$, generically lies in $GL(4,\IR)$, we have seen that  we can use an $SU(2) \times U(1)$ gauge transformation to write it in terms of a simpler matrix, $\hat m$,   in (\ref{DiagmP}).  This means that $m$ actually lies in $GL(2,\IC)$.  Similarly, the gauge fields are those of $SU(2) \times U(1)$ and the scalars, $\chi_I$, should be thought of as a complex doublet, $(\chi_1 + i \chi_2, \chi_3 - i \chi_4)$, transforming under these global and local symmetries.  Indeed, it might be natural to recast all the scalars in terms an element of $GL(3,\IC)$, or perhaps $SU(2,1)$.  Either way,   the BPS sector that we are studying  can be recast in terms of the unitary gauge group acting on complex fields with precise holomorphy properties.

\section{Supersymmetry in three dimensions}  
\label{sec:Supersymmetry}

We now use the results of the previous section to solve the three-dimensional BPS equations (\ref{fermsusy}).  We will, however, start with a summary of some of the conditions that will emerge from our computations.

\subsection{The supersymmetry conditions in three dimensions}
\label{sec:susycond}

We will  find that the superstrata preserve precisely the supersymmetries that satisfy:
\begin{equation}
    \Big(\oneone ~-~ \gamma^{12}\, \Gamma^{12}  \Big)\, \epsilon ~=~  0 \,, \qquad      \Big(\oneone ~+~ \gamma^{12}\, \Gamma^{34} \Big)\, \epsilon ~=~  0 \,.
    \label{basic-proj}
\end{equation}
Note that, together, these two projections imply:
\begin{equation}
    \Big(\oneone ~-~ \Gamma^{1234}  \Big)\, \epsilon ~=~  0 \,, 
    \label{int-proj}
\end{equation}
which was anticipated in (\ref{SO8proj2}). Moreover, any two of the  projectors from (\ref{basic-proj}) and (\ref{int-proj})  imply the third.

These projections mean that the superstratum  is a \nBPS{4} state in the $\Neql8$  theory of Section \ref{sec:Neql8}, or a  \nBPS{2} state in the $\Neql4$  theory of
Section \ref{sec:Neql4truncation}.  Either way, the superstratum has four residual supersymmetries, which is precisely consistent with its original formulation as a \nBPS{8} state in IIB supergravity.

Observe that the projection condition (\ref{int-proj}) implies that
\begin{equation}
\eta_j^{IJ} \,\Gamma^{IJ} \, \epsilon ~=~      0 \,,  \quad j=1,2,3 \,,
\label{sdproj}
\end{equation}
where the $\eta_j$ are defined in (\ref{etamat}).
This explicitly shows that the Killing spinor is a singlet under the $SU(2)$ generated by the $\eta_j$.  This means that under the  $SU(2) \times U(1)$  gauge transformation, $\mathscr{U}$, of (\ref{diagonalization}), the Killing spinor only transforms under the $U(1)$ part generated by $\bar\eta_3$.

This  also implies that the only part of $\cQ_\mu^{IJ}\Gamma^{IJ}$  that does not annihilate $\epsilon$ in  (\ref{Depsilon}), and hence in the BPS equations, (\ref{fermsusy}), are the pieces proportional to $\bar\eta_3$.

For the superstratum we take the momentum wave to be in the left-moving sector of the CFT, while the unbroken supersymmetry, and the unbroken $SU(2)$ $\cR$-symmetry,  act on the right-moving sector.  Since the supersymmetry is inert under the {\it local} $SU(2)$ gauge symmetry that we are using to simplify the background, we see that this  gauged $SU(2)$ must act purely on the left-moving sector.   This is in accord with the fact that this gauge symmetry is rearranging the fundamental momentum carrying modes that are encoded in the $\chi_i$.

For the $(1,0,n)$ superstratum, and for the general $(1,m,n)$ superstratum in the gauge discussed in Section  \ref{sec:gauge1mn}, and in which the scalar matrices take the form (\ref{DiagmP}), it is evident that we must take:
\begin{equation}
\mu_1 ~=~ \mu_2 ~=~ \log(\Lambda)  \,, \qquad \mu_3 ~=~ \mu_4 ~=~  0 \,, 
    \label{mures}
\end{equation}
in (\ref{Pdiag2}).

This constraint on the scalar matrix is a direct consequence of imposing:
\begin{equation}
\frac{\partial W}{\partial \mu_3} ~=~ \frac{\partial W}{\partial \mu_4} ~=~  0 \,, 
\label{partialWres}
\end{equation}
on the superpotential (\ref{superpot}) and then using (\ref{mures}). We will show how (\ref{partialWres}) emerges from solving the BPS equations.

Finally, we note that  (\ref{mures}) leads to
\begin{equation}
\frac{\partial W}{\partial \mu_1} ~=~ \frac{\partial W}{\partial \mu_2} ~=~  \frac{g_0}{2\,\Lambda^2} \, \big(1 - \Lambda^2\big) \,, \qquad \frac{\partial W}{\partial \chi_i} ~=~  -\frac{g_0}{4\,\Lambda^2}\, \chi_i  \,, \qquad W ~=~  -\frac{g_0}{2\,\Lambda^2}\,, 
\label{moreWres}
\end{equation}
and these relationships will also emerge from solving the BPS equations.

\section{Supersymmetry and the  $(1,0,n)$ superstratum}
\label{sec:10n_supersymmetry}

As a ``warm-up'' exercise we  start with the simpler, pure  $(1,0,n)$ superstratum, with  $\chi_3 =\chi_4 =0$, and whose key data is summarized in Section \ref{sec:gauge10n}. As we will see, the solution to the BPS equations for the general $(1,m,n)$ superstratum involves the solution of the $(1,0,n)$ system and so this ``exercise'' will, in fact, prove to be an essential part of the fully general system.

\subsection{The first BPS equation}
\label{sec:firstBPS}

The simplest equation to solve  is the fermion variation of (\ref{fermsusy}) for $r=5$:
\begin{equation}
\coeff{1}{2}\,i \,\Gamma^I_{A\dot{A}}\gamma^\mu\epsilon^A \, \cP_\mu^{I5}  ~-~  A_2^{A\dot{A} 5}\epsilon^A ~=~ 0\,.
\label{firstvar1}
\end{equation}
where we have set $g=-1$.  Using (\ref{u1xu1_scalar_kin}) and (\ref{A1A2tens}),   (\ref{mures}) and  (\ref{moreWres}), this becomes
\begin{equation}
i \,\Gamma^I_{A\dot{A}}\gamma^\mu\epsilon^A \, (\cD_\mu \chi)_I   ~+~ g_0 \, \sum_{j=1}^2 \, \chi_j \, (\Gamma^{1234}\,\Gamma^{j}\big)_{A \dot A} \epsilon^A ~=~ 0\,,
\label{firstvar2}
\end{equation}
where the sum is only over $j=1,2$ because $\chi_3 =\chi_4 =0$.

Next we use (\ref{AAbelform1}) and (\ref{holdiffchi}) to see that
\begin{equation}
\begin{aligned}
 \gamma^\mu  \, (\cD_\mu \chi)_1  ~=~ & (\dd \chi_1+  \cA \, \chi_2)_1\,\gamma^1  ~+~ (\dd \chi_1  +  \cA \, \chi_2)_2\, \gamma^2 ~-~g_0 \, \gamma^0 \, \chi_2\,, \\
  \gamma^\mu \, (\cD_\mu \chi)_2  ~=~ & (\dd \chi_2  -  \cA \, \chi_1 )_1\,\gamma^1  ~+~ (\dd \chi_2  -  \cA \, \chi_1 )_2\, \gamma^2~+~g_0 \, \gamma^0\, \chi_1 \,.
 \end{aligned}
\label{Dslashchi1}
\end{equation}
Now use the Cauchy-Riemann conditions  (\ref{chiCR}) and collect terms in  (\ref{firstvar2}):
\begin{equation}
\begin{aligned}
\Big[ \,& i \,  (\dd \chi_1+  \cA \, \chi_2)_1\, \Big(\gamma^1\, \Gamma^1_{A\dot{A}} +  \gamma^2\, \Gamma^2_{A\dot{A}}\Big) ~+~  i \,  (\dd \chi_1  +  \cA \, \chi_2)_2 \, \Big(\gamma^2\, \Gamma^1_{A\dot{A}} -  \gamma^1\, \Gamma^2_{A\dot{A}}\Big) \\
& ~+~ g_0\, \Big( - i \,  \gamma^0  \, \Gamma^1_{A\dot{A}} \, \chi_2  ~+~ i \,  \gamma^0  \, \Gamma^2_{A\dot{A}}  \, \chi_1 ~+~  (\Gamma^{1234}\,\Gamma^{1}\big)_{A \dot A}\, \chi_1~+~  (\Gamma^{1234}\,\Gamma^{2}\big)_{A \dot A}\, \chi_2\,\Big)  \, \Big]\,\epsilon^A ~ =~ 0\,.
 \end{aligned}
\label{firstvar3}
\end{equation}
Right multiply this expression by $\Gamma^2_{\dot A B}$,  use (\ref{st-gamas}) to replace $\gamma^0 = i \gamma^{12}$  and use the properties of the $\Gamma^I$'s  to write $\Gamma^{IJ}_{A B}  = -\Gamma^{IJ}_{BA} $ and  $\Gamma^{1234}_{A B} =\Gamma^{1234}_{BA} $.  This yields: 
\begin{equation}
\begin{aligned}
\Big[ \,& i \,  (\dd \chi_1+  \cA \, \chi_2)_1\, \gamma^2\,\big(-\gamma^{12}\, \Gamma^{12}_{BA} + \delta_{BA}\big) ~+~  i \,  (\dd \chi_1  +  \cA \, \chi_2)_2 \,\gamma^1\, \big(\gamma^{12}\, \Gamma^{12}_{BA} -    \delta_{BA} \big) \\
& ~-~ g_0\,  \gamma^{12} \,   \big(\Gamma^{12}_{BC} \,\big(\delta_{CA} ~+~   \gamma^{12} \,   \Gamma^{34}_{CA} \big)  \, \chi_2  ~+~\big(   \delta_{BA} ~+~ \gamma^{12}  \, \Gamma^{34}_{BA}\big)\,\chi_1\big)  \, \Big]\,\epsilon^A ~ =~ 0\,.
 \end{aligned}
\label{firstvar4}
\end{equation}
This vanishes by virtue of the projectors    (\ref{basic-proj}).

From this variation we see that holomorphy implies both the projection conditions in (\ref{basic-proj}) and, conversely, if one imposes these projectors first, the vanishing of the variation requires the Cauchy-Riemann conditions,  (\ref{chiCR}), and hence holomorphy.

\subsection{The second set of BPS equations}
\label{sec:secondBPS}

Using (\ref{fermsusy})  and (\ref{A1A2tens}), the next set of BPS equations may be written:
\begin{equation}
 \coeff{1}{2}\,i \,\Gamma^I_{A\dot{A}}\gamma^\mu\epsilon^A \, \cP_\mu^{Ir}  ~-~ \frac{\partial W}{\partial \mu_r} \, (\Gamma^{1234}\,\Gamma^{r}\big)_{A \dot A}   \,\epsilon^A ~=~ 0 \,, \qquad r=1,2,3,4 \,.
\label{secondvar1}
\end{equation}
From (\ref{u1xu1_cs}) one knows that $\cP_\mu^{Ir} = 0$ for  $r=3,4$,  and so this equation is satisfied by imposing the conditions (\ref{partialWres}).  The equation is non-trivial for $r=1,2$.

To simplify the equation we right multiply  by $\Gamma^r_{\dot A B}$ and then use $\Gamma^{IJ}_{A B}  = -\Gamma^{IJ}_{BA} $ and  $\Gamma^{1234}_{A B} =\Gamma^{1234}_{BA} $.  This yields the following equation for both $r=1$ and $r=2$: 
\begin{equation}
\begin{aligned}
\frac{i}{4\Lambda^2}\, \Big[\,\big( \lambda_1 \, \gamma^1 + \lambda_2  \, \gamma^2 \big)\,  \delta_{AB} ~-~  \Gamma^{12}_{BA}\, \big( 2 g_0 \,(1-\Lambda^2)\, \gamma^0 &  ~+~  \lambda_2 \,  \gamma^1 ~-~ \lambda_1 \, \gamma^2 \big) \,\Big]\, \epsilon^A \\
&  ~-~  \frac{g_0}{2\,\Lambda^2} \, \big(1 - \Lambda^2\big) \, (\Gamma^{1234}\big)_{BA}   \,\epsilon^A ~=~ 0 \,,
\end{aligned}
\label{secondvar2}
\end{equation}
where we have used (\ref{u1xu1_scalar_kin}) and (\ref{u1xu1_cs}) to replace $\cP^{Ir}$ and $\mathscr{C}^{12}$, and (\ref{moreWres}) to replace the derivative of $W$.

Using  (\ref{st-gamas}) to replace $\gamma^0 = i \gamma^{12}$  and collecting terms, one finds:
\begin{equation}
\frac{i}{4\Lambda^2}\, \Big[\,\big( \lambda_1 \, \gamma^1 + \lambda_2  \, \gamma^2 \big)\,  \big( \delta_{AB}   ~-~\gamma^{12}\, \Gamma^{12}_{BA}\, \big)   \,\Big]\, \epsilon^A  ~+~  \frac{g_0}{2\,\Lambda^2} \, \big(1 - \Lambda^2\big) \, \Gamma^{12}_{BC}\, \big(\gamma^{12}  \, \delta_{CA}  ~-~ \Gamma^{34}_{CA}\big)   \,\epsilon^A ~=~ 0 \,,
\label{secondvar3}
\end{equation}
and the left-hand side vanishes by virtue of the projectors  (\ref{basic-proj}).

In this BPS equation, the derivatives of $\Lambda$ are cancelled by the Chern-Simons connection, and the superpotential term is cancelled by the time-like component of the gauge connection.

\subsection{The third  BPS equation}
\label{sec:thirdBPS}

This is the gravitino variation from (\ref{fermsusy}):
\begin{equation}
D_\mu \epsilon^A ~-~ i A_1^{AB}\gamma_\mu\epsilon^B ~=~ 0 \,.
\label{thirdvar1}
\end{equation}
Writing out all the pieces more explicitly and using (\ref{Depsilon}) and (\ref{A1A2tens}), and converting to forms gives
\begin{equation}
\dd \epsilon^A    ~+~ \coeff{1}{4}\,\omega_{ab}\,\gamma^{ab}\, \epsilon^A + \coeff{1}{4}\,\cQ^{IJ}\Gamma^{IJ}_{AB}\, \epsilon^B  ~-~ i W\, (\Gamma^{1234}\big)_{AB} \,\gamma_c \, e^c \,\epsilon^B ~=~ 0 \,.
\label{thirdvar2}
\end{equation}
Now use (\ref{Qsimp1}), (\ref{u1xu1_gauge}), (\ref{u1xu1_cs}), (\ref{spin_connection}) and (\ref{moreWres}) to replace all the pieces:
\begin{equation}
\begin{aligned}
\dd \epsilon^A   & ~+~ \bigg[\, \frac{g_0}{2 \Lambda^2} \,  \big(   \gamma^{01} \,e^2 - \gamma^{02} \,e^1   -  \gamma^{12} \,e^0 \big) - \frac{1}{4 \Lambda^2}\,\big( \lambda_2 \, e^1- \lambda_1\,e^2 \big)\,\gamma^{12} + \frac{g_0}{2\Lambda}  \,\frac{2 \,r^2 + a^2}{r\, \sqrt{r^2 + a^2}}  \,e^2\,\gamma^{12} \, \bigg]\,\epsilon^A\\ 
&  ~-~\bigg[\, \frac{g_0}{2} \, \Big( e^0 - \frac{r}{\sqrt{r^2 + a^2}} \frac{e^2}{\Lambda} \Big)  ~+~ \frac{1}{4\Lambda^2} \,\Big( 2 g_0 \,(1-\Lambda^2)\, e^0 + \big(  \lambda_2 \,  e^1 - \lambda_1 \, e^2 \big)\,\Big) \,\bigg] \,\Gamma^{12}_{AB}\, \epsilon^B  \\
&~+~ g_0 \,\bigg[\,  \frac{e^0}{2 \Lambda^2} - \frac{r}{\sqrt{r^2 + a^2}} \, \frac{e^2}{2\Lambda}\,\bigg] \,\Gamma^{34}_{AB}\, \epsilon^B   \\
&~+~ i   \frac{g_0}{2\,\Lambda^2}\, \big(\gamma^0 \, e^0  - \gamma^1 \, e^1 -\gamma^2 \, e^2\big) \,(\Gamma^{1234}\big)_{AB} \,\epsilon^B ~=~ 0 \,.
\end{aligned}
\label{thirdvar3}
\end{equation}
Note that one has to be very careful about raising and lowering indices correctly using the metric with $(+ - -)$ signature. 

Collecting terms and replacing $\gamma^0 = i \gamma^{12}$, we find
\begin{equation}
\begin{aligned}
\dd \epsilon  & ~+~ \frac{i g_0}{2 \Lambda^2} \,  \big(   \gamma^{1} \,e^1 + \gamma^{2} \,e^2   \big)\big(1 -\Gamma^{1234}\big)\, \epsilon - \frac{1}{4 \Lambda^2}\,\big( \lambda_2 \, e^1- \lambda_1\,e^2 \big)\,\big( \gamma^{12} + \,\Gamma^{12}\, \big)  \,\epsilon \\ 
&  ~-~  \frac{ g_0}{2\Lambda^2} \,   e^0\,\big(\Gamma^{12}- \Gamma^{34}+ \gamma^{12} + \gamma^{12} \, \Gamma^{1234} \big )  \, \epsilon  \\
&~+~ \frac{g_0}{2\Lambda}\,e^2\,\frac{1}{r \, \sqrt{r^2 + a^2}} \, \bigg[\, (2 \,r^2 + a^2)\,\gamma^{12} ~+~ r^2  \,\Gamma^{12}   ~-~   r^2 \, \Gamma^{34}\,\bigg] \, \epsilon   ~=~ 0  \,.
\end{aligned}
\label{thirdvar4}
\end{equation}
Almost all the terms vanish as a result of the projectors (\ref{basic-proj}) and (\ref{int-proj}), leaving the equation:
\begin{equation}
\dd \epsilon   ~+~  \frac{g_0}{2\Lambda}\,\frac{a^2 }{r \, \sqrt{r^2 + a^2}} \,e^2 \,\gamma^{12}\, \epsilon   ~=~ 0  \qquad \Leftrightarrow \qquad \dd \epsilon   ~+~  \frac{1}{\sqrt{2}  \, R_y} \,\gamma^{12}\, \dd v \, \epsilon   ~=~ 0\,.
\label{thirdvar5}
\end{equation}
This has an elementary solution: 
\begin{equation}
\epsilon   ~=~ \exp \bigg[\,  \frac{v}{\sqrt{2}  \, R_y} \,\gamma^{12} \, \bigg] \, \epsilon_0 ~=~ \exp \bigg[\, - \frac{v}{\sqrt{2}  \, R_y} \,\Gamma^{12} \, \bigg] \, \epsilon_0 ~=~ \exp \bigg[\,  \frac{v}{\sqrt{2}  \, R_y} \,\Gamma^{34} \, \bigg] \, \epsilon_0 \,,
\label{epssol1}
\end{equation}
where $\epsilon_0$ is a constant spinor satisfying the projection conditions (\ref{basic-proj}) and (\ref{int-proj}). 

Interestingly enough, the last rotation in (\ref{epssol1}) is induced by the gauge rotation we did in Section \ref{sec:ssscalars} to re-write the scalar fields, $\chi$,  of 
\cite{Mayerson:2020tcl}  in the form (\ref{chi_superstrata}).  This means that, had we stayed in the original gauge of \cite{Mayerson:2020tcl}, the cancellation in (\ref{thirdvar4}) would have been complete, and the Killing spinor, $\epsilon$, would have been constant.  In retrospect, this is not surprising because the three-dimensional fields obtained in  \cite{Mayerson:2020tcl} descend from the six-dimensional solution, and in the  six-dimensional theory the supersymmetry parameter is constant. (See (2.28) of \cite{Gutowski:2003rg}.)

There are some interesting features in this computation.  First, we note that the spin-connection terms involving $d \Lambda$ are cancelled by the Chern-Simons connection, and so the existence of the supersymmetry hinges on the ``twisting of the spin-connection'' by the Chern-Simons connection.  It is also evident from the computation that the cancellation requires that the metric have the K\"ahler fibration form of the metric discussed in Section \ref{sec:thessmet}, but all we needed to know about the two-dimensional spatial base was that it had constant negative curvature.  Thus one could use a uniformized Riemann surface instead of the hyperboloid of (\ref{2dbasemet}).  Using a Riemann surface would generically lead to strange boundaries at infinity, but there still may be interesting BPS solutions based on such a generalization.

\section{Supersymmetry in the generic  $(1,m,n)$ superstratum}
\label{sec:1mn_supersymmetry}

We now solve the BPS equations for a generic $(1,m,n)$ superstratum. When working in the gauge defined in section \ref{sec:gauge1mn}, one finds that each equation decomposed into a sum of the terms found in the pure $(1,0,n)$ superstratum, plus additional terms. Since the first ones have been shown to vanish in Section \ref{sec:10n_supersymmetry}, our goal here is to show that the additional terms also vanish when one applies the projection conditions.

\subsection{The first BPS equation}
\label{sub:the_first_bps_equation_1mn}

We start with the first BPS equation (\ref{firstvar1}), the fermion variation for $r=5$ , that we recall here:
\begin{equation}
\coeff{1}{2}\,i \,\Gamma^I_{A\dot{A}}\gamma^\mu\epsilon^A \, \cP_\mu^{I5}  ~-~  A_2^{A\dot{A} 5}\epsilon^A ~=~ 0\,.
\end{equation}

Using (\ref{genscalar_kin}), (\ref{A1A2tens}),  (\ref{trfchi}) and (\ref{tildechi}), we find
\begin{equation}
i \,\Gamma^I_{A\dot{A}}\gamma^\mu\epsilon^A \, \bigg(\cD_\mu \bigg(\frac{\rho_0}{\rho_1}\, \tilde \chi\bigg)\bigg)_I   ~+~ g_0 \, \sum_{j=1}^2 \, \frac{\rho_0}{\rho_1} \, \chi_j \, (\Gamma^{1234}\,\Gamma^{j}\big)_{A \dot A} \epsilon^A ~=~ 0\,,
\label{firstvar1mn}
\end{equation}

Next, using (\ref{Ahatdefn}) and (\ref{AAbelparts}), we compute the covariant derivative
\begin{equation}
\bigg(\cD_\mu \bigg(\frac{\rho_0}{\rho_1}\,\tilde \chi\bigg)\bigg)_I ~=~ \frac{\rho_0}{\rho_1}\, (\hat\cD_\mu \tilde \chi)_I +  \bigg(\partial_\mu \frac{\rho_0}{\rho_1} \bigg) \,\tilde\chi_I - \frac{\rho_0}{\rho_1} \, \qty((\partial_\mu{\mathscr{U}}) \mathscr{U}^{-1} \tilde \chi)_I \ ,
\end{equation}
where $\hat\cD$ denotes the covariant derivative of the abelian gauge field :
\begin{equation}
    (\hat\cD_\mu \tilde \chi)_I ~=~ \partial_\mu \chi_I - 2 g_0 (\tilde A_\text{Abelian} \tilde\chi)_I
\end{equation}

This implies that the BPS equation takes the form
\begin{equation}
\begin{aligned}
    \frac{\rho_0}{\rho_1} \Big(i \,\Gamma^I_{A\dot{A}}\gamma^\mu\epsilon^A \, (\hat\cD_\mu \tilde \chi)_I   &~+~ g_0 \, \sum_{j=1}^2 \, \chi_j \, (\Gamma^{1234}\,\Gamma^{j}\big)_{A \dot A} \epsilon^A\Big) \\
    &+ i \,\Gamma^I_{A\dot{A}}\gamma^\mu\epsilon^A \qty( \qty(\partial_\mu \frac{\rho_0}{\rho_1}) \tilde \chi_I ~- ~ \frac{\rho_0}{\rho_1} \qty((\partial_\mu {\mathscr{U}}) \mathscr{U}^{-1}\tilde  \chi)_I ) ~=~ 0 \ .
\end{aligned}
\end{equation}
The first line of this equaton is simply a multiple of equation (\ref{firstvar2}) and the analysis in Section \ref{sec:firstBPS} showed that it vanishes.

To simplify the second term, we use (\ref{puregauge}), (\ref{rho_defs}), (\ref{KLdefn}), (\ref{tildechi}), (\ref{etamat}) and (\ref{etabarmat}).  The simple form of $\tilde \chi$, (\ref{tildechi}), picks out specific matrix elements of $\eta$ and $\bar \eta$  in (\ref{puregauge}), and the result is:
\begin{equation}
\begin{aligned}
i \,  \Big[ &-\frac{\rho_2^2}{\rho_0 \rho_1} (L_1-L_2)_\mu \gamma^\mu \qty(\chi_1 \Gamma^1_{A\dot{A}} + \chi_2 \Gamma^2_{A\dot{A}}) ~+~ \frac{\rho_2^2}{\rho_0\rho_1} (K_1 + K_2)_\mu \gamma^\mu \qty(\chi_1 \Gamma^2_{A\dot{A}} - \chi_2 \Gamma^1_{A\dot{A}})
    \\
    &+~ \frac{\rho_2}{\rho_0} (L_1 - L_2)_\mu \gamma^\mu \qty(\chi_1 \Gamma^4_{A\dot{A}} + \chi_2 \Gamma^3_{A\dot{A}})  ~-~ \frac{\rho_2}{\rho_0} (K_1 + K_2)_\mu \gamma^\mu \qty(\chi_1 \Gamma^3_{A\dot{A}} - \chi_2 \Gamma^4_{A\dot{A}}) \Big] \epsilon^A ~=~ 0 \ .
\end{aligned}
\end{equation}

Observe that  the interchange:
\begin{equation}
     \Gamma_1 \leftrightarrow \Gamma_4 \ , \qquad \Gamma_2 \leftrightarrow \Gamma_3 \ ,
\end{equation}
maps the first line of the equation onto the second, up to an overall factor, and vice-versa. Since the projectors (\ref{basic-proj}) are invariant under this transformation, it therefore suffices to show that the first line of the equation vanishes. 

Now multiply the first line of the equation by $\Gamma^1_{\dot{A}B}$, use the notation  (\ref{cKcLdefn}) and the consequences of holomorphy (\ref{relationKL}) to reduce it to
\begin{equation}
\begin{aligned}
 - i \,     \frac{\rho_2^2}{\rho_0 \rho_1} \Big[ &  \mathcal{L}_1 \chi_1 \qty(\gamma^1 \delta_{AB} +  \gamma^2  \Gamma^{21}_{AB})\epsilon^A ~+~ \mathcal{L}_1 \chi_2 \qty(\gamma^1   \Gamma^{21}_{AB} - \gamma^2 \delta_{AB}) \epsilon^A \\
    &+~ \mathcal{L}_2 \chi_1 \qty( \gamma^2 \delta_{AB} - \gamma^1   \Gamma^{21}_{AB})\epsilon^A ~+~ \mathcal{L}_2 \chi_2 \qty(\gamma^2  \Gamma^{21}_{AB}+ \gamma^1 \delta_{AB}) \epsilon^A \Big] \ .
\end{aligned}
\end{equation}
Using the fact that  $\Gamma^{21}_{AB}  = + \Gamma^{12}_{BA}$, one can sees that each of these terms vanishes due to the projections (\ref{basic-proj}). Once again, we see that holomorphy and the projectors are working together to produce the residual supersymmetry.  

\subsection{The second set of BPS equations}
\label{sec:secondBPS1mn}

The next set of BPS equation is given by (\ref{secondvar1}):
\begin{equation}
 \coeff{1}{2}\,i \,\Gamma^I_{A\dot{A}}\gamma^\mu\epsilon^A \, \cP_\mu^{Ir}  ~-~ \frac{\partial W}{\partial \mu_r} \, (\Gamma^{1234}\,\Gamma^{r}\big)_{A \dot A}   \,\epsilon^A ~=~ 0 \,, \qquad r=1,2,3,4 \,.
 \label{secondvar1mn}
\end{equation}
where the scalar kinetic terms are now given by (\ref{PIJfinal}).

First we note that because of the cancellations between the gauge connection and the Chern-Simons connection discussed in Section \ref{sec:holomorphy}, we have that 
$ \cP_\mu^{Ir} = 0$ for   $r=3,4$.  (See (\ref{PIJfinal}).)  This means that we must also impose (\ref{partialWres}) and then, with this condition, the BPS equation  (\ref{secondvar1mn}) is trivially satisfied for $r=3,4$.

For $r=1,2$, using the expression of the Chern-Simons term (\ref{fullCSterm2}), the equation has the form
\begin{equation}
\begin{aligned}
    \coeff{1}{2}\,i \,\Gamma^I_{A\dot{A}}\gamma^\mu\epsilon^A \, &\Big(\dd{\log(\Lambda)} \delta^{Ir} - \coeff{1}{2} \mathscr{C}^{Ir}_\text{Abelian}\Big)  ~-~ \frac{\partial W}{\partial \mu_r} \, (\Gamma^{1234}\,\Gamma^{r}\big)_{A \dot A} \,\epsilon^A \\
    &-~ i\,\frac{\rho_1\rho_2}{8\Lambda}\Big[(L_1-L_2)_\mu \gamma^\mu \eta_1^{Ir} + (K_1 + K_2)_\mu \gamma^\mu \eta_2^{Ir}\Big] \Gamma^I_{A\dot{A}}\epsilon^A ~=~ 0 \ ,
\end{aligned}
\end{equation}
where $\mathscr{C}^{IJ}_\text{Abelian}$ is the Abelian Chern-Simons term  in (\ref{CAbelian}) and (\ref{u1xu1_cs}).

It was shown in Section \ref{sec:secondBPS} that the first line of this equation vanishes. We multiply what remains by $\Gamma^{r+2}_{\dot{A}B}$ and then use $\Gamma^{IJ}_{A B}  = -\Gamma^{IJ}_{BA}$ to find for both $r=1$ and $r=2$
\begin{equation}
    i\,\frac{\rho_1\rho_2}{8\Lambda} \Big[ (L_1 - L_2)_\mu \gamma^\mu \delta_{AB} + (K_1 + K_2)_\mu \gamma^\mu \Gamma^{34}_{BA} \Big] \epsilon^A \,.
\end{equation}

We use once again the relation between the derivatives resulting from the holomorphy (\ref{relationKL}) to obtain:
\begin{equation}
    i\,\frac{\rho_1\rho_2}{8 \Lambda} \Big[ \mathcal{L}_1 \qty(\gamma^1 \delta_{AB} - \gamma^2 \Gamma^{34}_{AB}) + \mathcal{L}_2 \qty(\gamma^2 \delta_{AB} + \gamma^1 \Gamma^{34}_{BA})  \Big] \epsilon^A \,.
\end{equation}
This vanishes after using the projection conditions (\ref{basic-proj}).

\subsection{The third BPS equations}
\label{sec:thirdBPS1mn}

The third BPS equation is given by (\ref{thirdvar2}) :
\begin{equation}
\dd \epsilon^A    ~+~ \coeff{1}{4}\,\omega_{ab}\,\gamma^{ab}\, \epsilon^A + \coeff{1}{4}\,\cQ^{IJ}\Gamma^{IJ}_{AB}\, \epsilon^B  ~-~ i W\, (\Gamma^{1234}\big)_{AB} \,\gamma_c \, e^c \,\epsilon^B ~=~ 0 \,.
\end{equation}

Here, the spin connections and superpotential terms are the same as for the pure $(1,0,n)$ superstrata. The only difference arises from the connection, $\cQ^{IJ}$, given by (\ref{QIJfinal}). While   generically an $SO(4)$ connection,  the connection, $\cQ^{IJ}$, of (\ref{QIJfinal}) lives in $SU(2)\times U(1)$ and, as noted in Section \ref{sec:susycond}, the projection condition (\ref{int-proj}) implies that the Killing spinor is a singlet under $SU(2)$.  In particular, the spinor satisfies (\ref{sdproj}). This tremendously simplifies  the computation: one can ignore all the terms in $\eta_1$, $\eta_2$ or $\eta_3$ in $\cQ$, and keep only the terms in $\bar\eta_3$.

Using the result (\ref{QIJfinal}) with (\ref{Ahatdefn}) and   (\ref{puregauge}), one obtains
\begin{equation}
    \coeff{1}{4} \cQ^{IJ}\Gamma^{IJ}_{AB}\, \epsilon^B ~=~ \coeff{1}{4} \cQ^{IJ}_\text{Abelian}\Gamma^{IJ}_{AB}\, \epsilon^B ~-~ \coeff{1}{4} (K_1 + K_2) \qty(\Gamma^{12}_{AB} - \Gamma^{34}_{AB}) \epsilon^B \ ,
\end{equation}
where $\cQ^{IJ}_\text{Abelian}$ is the connection of the pure $(1,0,n)$ superstrata (\ref{Qsimp1}).

The full equation is then given by
\begin{equation}
\begin{aligned}
    \dd \epsilon^A   ~&-~ \coeff{1}{4} (K_1 + K_2) \,\qty(\Gamma^{12}_{AB} - \Gamma^{34}_{AB})\, \epsilon^B \\[1ex]
    &+~ \coeff{1}{4}\,\omega_{ab}\,\gamma^{ab}\, \epsilon^A + \coeff{1}{4}\,\cQ^{IJ}_\text{Abelian}\Gamma^{IJ}_{AB}\, \epsilon^B  ~-~ i W\, (\Gamma^{1234}\big)_{AB} \,\gamma_c \, e^c \,\epsilon^B ~=~ 0 \,.
\end{aligned}
\end{equation}

The second line of this equation is, once again, precisely the same as that of the $(1,0,n)$ superstratum, and, in Section \ref{sec:thirdBPS}, it was shown to reduce to the simple gauge term in (\ref{thirdvar5}).  Using the projectors (\ref{basic-proj}) one obtains:
\begin{equation}
    \dd \epsilon ~-~ \coeff{1}{2} (K_1 + K_2) \,\Gamma^{12}\, \epsilon ~-~  \frac{\dd v}{\sqrt{2}  \, R_y} \,\Gamma^{12}\,\epsilon ~=~ 0 \ .
    \label{diff_eq_1mn}
\end{equation}

The key to solving this differential equation is to use the expressions in (\ref{KLdefn}) to observe that  $-\coeff{1}{2}K_1$ and $\coeff{1}{2}K_2$ are the exterior derivatives of the phases of $\chi_1 + i \chi_2$ and, respectively, $\chi_3 - i \chi_4$. Introduce the phases $\phi_1$ and $\phi_2$ by defining
\begin{align}
    \chi_1 + i \chi_2 ~&\equiv~ \rho_1 \, e^{i \phi_1} \\
    \chi_3 - i \chi_4 ~&\equiv~ \rho_2 \, e^{i \phi_2}\ ,
\end{align}
then the equation becomes
\begin{equation}
    \dd\epsilon + \dd(\phi_1 - \phi_2 - \frac{v}{\sqrt{2}R_y}) \Gamma^{12} \epsilon ~=~ 0\ .
\end{equation}

A solution to this equation is
\begin{equation}
    \epsilon   ~=~ \exp \bigg[\, \qty(- \phi_1 + \phi_2 + \frac{v}{\sqrt{2}  \, R_y}) \,\Gamma^{12} \, \bigg] \, \epsilon_0 \ ,
 \label{epssol2}
\end{equation}
where $\epsilon_0$ is a constant spinor satisfying the projection conditions (\ref{basic-proj}) and (\ref{int-proj}).

Once again we could have anticipated this result from the six-dimensional formulation, where the Killing spinor is independent of all coordinates.  We have transformed that result first by the gauge transformation described in Section \ref{sec:ssscalars} to re-write the scalar fields, $\chi$,  of  \cite{Mayerson:2020tcl}  in the form (\ref{chi_superstrata}). In Section \ref{sec:thirdBPS} we described how this led to the explicit $v$-dependence in (\ref{epssol1}).  This is also the origin the explicit $v$-dependence in (\ref{epssol2}).  We have also   transformed the reduction of the six-dimensional  solution by  $\mathscr{U}$,  defined in (\ref{diagonalization}).  As we have noted, the Killing spinor only transforms under the $U(1)$ piece of this gauge transformation, and from  (\ref{puregauge}) we see that this $U(1)$ piece is:
\begin{equation}
\big(\dd \mathscr{U}_0   \big) \,\mathscr{U}_0^{-1} ~=~ \coeff{1}{2} \, (K_1 + K_2) \, \bar\eta_3   \,.
\label{puregaugeU1}
\end{equation}
It is precisely this term that leads the non-trivial $\chi$-dependent rotation in  (\ref{diff_eq_1mn}) and thus generates the non-trivial $\chi$-dependent behavior in the supersymmetry of  (\ref{epssol2}). 

We have now shown that the three-dimensional data for the $(1,m,n)$ superstrata \cite{Mayerson:2020tcl}  does indeed lead to a BPS solution, with four supersymmetries, in the  gauged three-dimensional supergravity described in Section \ref{Sect:3Dtheory}. 

\section{Final comments}
\label{sec:Conclusions}

In this paper we have taken the superstratum data from  \cite{Mayerson:2020tcl} and shown precisely how it leads to  a supersymmetric/BPS solution  in three dimensions.  The result is a smoothly-capped BTZ background with scalar and gauge field excitations that, together,  preserve four supersymmetries.  

Given the superstratum solutions in six dimensions, the existence of such a three-dimensional superstratum is not altogether surprising.   What is surprising is the central result of \cite{Mayerson:2020tcl} in which it was shown that there are families of superstrata that live in consistent, three-dimensional truncations. Here we have completed this  analysis  by showing how the BPS structure also survives the consistent truncation. Our computations are sensitive to multitudinous conventions, normalizations and signs, and the fact that the three-dimensional BPS equations are consistent with the six-dimensional BPS solutions is a testament to the accuracy of \cite{Mayerson:2020tcl},  the accuracy of the earlier work on three-dimensional gauged supergravities, and to our translation between these somewhat different formulations of the core underlying supergravity. 

Our three-dimensional analysis raises some interesting questions for the holographic theory and for the six-dimensional supergravity.  In solving the three-dimensional BPS equations, holomorphy and the Chern-Simons interaction played a crucial role.  It would be interesting to understand how this plays out in the dual CFT. At a more technical level,  we made heavy use of the {\it local}  $SU(2) \times U(1)$ gauge transformation {diagonalization} to convert the $(1,m,n)$ superstratum data into something closer to the $(1,0,n)$ superstratum data.  The CFT has an $SU(2)_L \times SU(2)_R$  current algebra and it would be very interesting to understand how these local gauge transformations are realized within these current algebras. 

Then there is the question of  the dimensional reduction of the fermions:  how do the three-dimensional fermions encode the six-dimensional fermions?   How are the $\cR$-symmetries and global symmetries in three and six dimensional supergravity  related to one another, and to the  $\cR$-symmetries and global symmetries of the CFT?  While the supersymmetry projectors   (\ref{basic-proj}) are very simple, and seem very canonical, one should be able to derive them from the full consistent truncation and knowledge of the CFT states.
 
It would also be very interesting to see if there are more general BPS solutions in three-dimensions. In Section \ref{sec:thirdBPS}  we already noted that there are  easy generalizations because the spatial metric, $ ds_2^2$, in (\ref{2dbasemet}) could be replaced by a constant (negative) curvature metric on a  Riemann surface of genus greater than one.   This would add some very interesting topology to the core of the solution, but at the cost of having some strange boundary conditions at infinity.  More broadly, our presentation has been very much driven by the knowledge of the data that comes from the superstrata.  It is possible that a more detailed analysis could lead to extensions, or generalizations, of these superstrata solutions.  Indeed, perturbative analysis in six-dimensions reveals that there are a lot more metric and flux modes that preserve supersymmetry and that are not part of the known superstratum solutions.  It remains to be seen whether any of these six-dimensional modes survive  in the consistent truncation to three dimensions, but if they do, one will then be able to construct new BPS solutions in three dimensions and then uplift them to create new superstrata in six dimensions. 

The remarkable thing about \cite{Mayerson:2020tcl}, and this follow-up, is that a really non-trivial sector of the D1-D5-P system, with a known holographic dual, can be reduced to three-dimensional supergravity.  As we remarked in the introduction, we hope to use this observation to obtain non-extremal superstrata.  While we are very optimistic about finding new non-extremal solutions in this manner, we should also add some words of caution.  Consistent truncations are very much a ``Faustian Bargain'': to obtain a much simpler, closed set of equations one must truncate away many of the higher-dimensional degrees of freedom.  This means that the lower dimensional, gauged supergravity may  lack the fidelity needed to describe the correct physics.  There are many examples of this in holographic RG flows, such as trying to describe a confining RG flow in $\Neql4$ Yang-Mills theory using gauged, five-dimensional supergravity \cite{Girardello:1999bd,Freedman:1999gp,Pilch:2000fu, Petrini:2018pjk, Bobev:2018eer,Bena:2018vtu}.  The failure of the gauged supergravity to capture the physics typically results in a singular, unphysical solution. The correct physics can only be described if one activates all the essential degrees of freedom needed to describe that physics, and this may require the higher-dimensional supergravity, or even the full string theory.  

Indeed this issue lies at the heart of the microstate geometry and fuzzball programs:  General Relativity in four dimensions  captures the large-scale gravitational aspects of black-hole physics but it requires string theory, or, at least, supergravity, to describe the information that is apparently ``lost''   by  four-dimensional General Relativity.   Thus singularities, and horizons, are to be thought of as a pathology brought on by restricting the degrees of freedom to a theory that does not have the proper fidelity to resolve the physics. These dangers are inherent in consistent truncations, and so it is possible that three-dimensional supergravity might not have the degrees of freedom necessary to describe non-extremal microstate structure. It is, of course, becoming increasingly evident that matter falling into microstate geometries ultimately scrambles into stringy excitations around those geometries (see, for example,  \cite{Marolf:2016nwu,Martinec:2020cml}), and so not even supergravity is capable of describing this microstate structure.  On the other hand, it is very plausible that supergravity will capture some coherent families of non-extremal microstate geometries and, if one is optimistic, it is also possible that some of them will be captured by three-dimensional gauged supergravity. We are indeed optimistic that this will occur, but we note that if the non-extremal solutions of three-dimensional supergravity all turn out to be singular, this will almost certainly represent a lack of fidelity in the three-dimensional theory rather than a problem with the microstate geometry and fuzzball programmes.  Resolving this issue is for the future.

Returning to the results presented in this paper, and  putting the considerations of superstrata, consistent truncation and black-hole microstructure to one side, it is important to note that, from the three-dimensional perspective, the BPS solutions we describe here are completely new, smooth, three-dimensional backgrounds, and are thus intrinsically interesting in their own right.

\vspace{0.8cm}

\section*{Acknowledgments}
\vspace{-2mm}
We would like to thank  Daniel Mayerson and, especially, Robert Walker for discussions in the early stages of this project.  This work was supported, in part, by ERC Grant number: 787320 - QBH Structure and by DOE grant DE-SC0011687. 

\appendix

\section{Basis changes}

\subsection{$SO(4,5)$ basis changes}

We use both the canonical and $GL(4\,,\IR)$ bases for  $SO(4,5)$.  In these bases the invariant metrics are, respectively: 
\begin{equation}
\eta 
~\equiv~
\left( \begin{matrix} 
 \oneone_{4 \times 4} &0_{4 \times 5}\\
0_{5 \times 4} & -\oneone_{5 \times 5} 
\end{matrix} \right)  \,,
\qquad 
\hat \eta ~\equiv~
\left( \begin{matrix} 
0_{4 \times 4} & \oneone_{4 \times 4} & 0 \\
\oneone_{4 \times 4} & 0_{4 \times 4} & 0 \\
0 &0& -1
\end{matrix} \right) \,.
\label{invmat1a}
\end{equation}
The change of basis matrix, $\mathscr{B}$, is thus required to satisfy: 
\begin{equation}
\eta ~=~ \mathscr{B} \, \hat \eta \, \mathscr{B}^{-1}
\label{cofbasis}
\end{equation}
This determines $\mathscr{B}$ up to some signs.  We use: 
\begin{equation}
\mathscr{B}
~\equiv~
\left( \begin{matrix} 
\frac{1}{\sqrt{2}}\, \oneone_{4 \times 4} & \frac{1}{\sqrt{2}}\, \oneone_{4 \times 4} & 0 \\
- \frac{1}{\sqrt{2}}\, \oneone_{4 \times 4} & \frac{1}{\sqrt{2}}\, \oneone_{4 \times 4} & 0 \\
0 &0& 1
\end{matrix} \right) \,,
\label{cofbasismat}
\end{equation}
which is an orthogonal matrix with determinant equal to one.

\newpage

\begin{adjustwidth}{-1mm}{-1mm} 

\bibliographystyle{utphys}      

\bibliography{microstates}       

\end{adjustwidth}

\end{document}